\documentclass[aps,pra,reprint,groupedaddress]{revtex4-2}

\usepackage[english]{babel}
\usepackage{graphics}
\usepackage{todonotes}
\usepackage{pstricks}
\usepackage{graphicx}

\usepackage{color}
\usepackage{braket}
\usepackage{booktabs} 
\usepackage{mathtools}
\definecolor{darkblue}{rgb}{0.0,0.0,0.4}
\definecolor{darkgreen}{rgb}{0.0,0.4,0.0}
\definecolor{darkred}{rgb}{0.6,0.0,0.0}
\usepackage[bookmarksopen]{hyperref}
\hypersetup{colorlinks,linkcolor=darkblue,citecolor=darkblue,urlcolor=darkblue}

\newcommand{\Op}[1]{\mathbf{#1}} 
\newcommand{\Rsc}{R_{\mathrm{sc}}} 
\newcommand{\baf}{{${}^{137}$BaF}}
\newcommand{\bafnineteen}{{${}^{137}\mathrm{Ba}^{19}\mathrm{F}$}}
\newcommand{\baff}{{${}^{138}$BaF}}
\newcommand{\gs}{{X${}^{2}\Sigma^{+}$}}
\newcommand{\exs}{{A${}^{2}\Pi_{1/2}$}}
\newcommand{\Wj}[6]{ \begin{pmatrix}
  #1 & #2 & #3 \\
  #4 & #5 & #6
 \end{pmatrix}}

\begin{document}

\title{A laser cooling scheme for precision measurements\\using barium monofluoride (\bafnineteen) molecules}

\author{Felix Kogel}
\author{Marian Rockenh\"auser}
\author{Ralf Albrecht}

\author{Tim Langen}
\email{t.langen@physik.uni-stuttgart.de}

\affiliation{5. Physikalisches  Institut  and  Center  for  Integrated  Quantum  Science  and  Technology,Universit\"at  Stuttgart,  Pfaffenwaldring  57,  70569  Stuttgart,  Germany}

\begin{abstract}
We theoretically investigate the laser cooling of fermionic barium monofluoride (\baf) molecules, which are  promising candidates for precision studies of weak parity violation and nuclear anapole moments. This molecular species features two nuclear spins, resulting in a hyperfine structure that is considerably more complicated than the one found in the usual laser-cooled diatomics. We use optical Bloch equations and rate equations to show that optical cycling, sub-Doppler cooling and bichromatic forces can all be realized under realistically achievable experimental conditions. 
\end{abstract}

\maketitle

\section{Introduction}
Cold and ultracold molecules promise a large variety of applications ranging from precision measurements to quantum chemistry to many-body physics~\cite{Carr2009,Bohn2017,DeMille2017,Safronova2018}. This has recently triggered important progress in their direct cooling and manipulation using laser cooling techniques~\cite{Fitch2021}. Several diatomic species have been trapped and cooled to the microkelvin regime~\cite{Barry2014,Truppe2017,Anderegg2017,Collopy2018,Anderegg2018,McCarron2018,Ding2020}, where they can be coherently manipulated~\cite{Williams2018} and individually addressed~\cite{Anderegg2019,Cheuk2020}. There is an ongoing effort to extend these powerful techniques to many more molecular species~\cite{Chen2017,Norrgard2017,Lim2018,Truppe2019,McNally2020,Albrecht2020}, including polyatomics~\cite{Isaev2016,Kozyryev2017,Baum2020,Mitra2020}. 

Here, we investigate a laser cooling scheme for the \baf\, isotopologue of barium monofluoride (BaF). This molecular species is a promising candidate for studies of nuclear spin-dependent parity violation, which aim to precisely determine anapole moments and other important consequences of the weak interaction~\cite{Safronova2018,Wood1997,Demille2008,Altuntas2018}. Other isotopologues of BaF have also been considered for precision measurements of the electron's permanent electric dipole moment~\citep{Aggarwal2018}.  For such precision experiments, laser cooling and optical cycling promise better statistics, more efficient detection and thus higher sensitivity to the observables of interest. 

However, \baf\, features two nuclear spins, which leads to a complex hyperfine structure with many dozens of resolved levels that render laser cooling challenging. In the following, we demonstrate that laser cooling can still be possible under realistic conditions. 

While we take \baf\, as a specific example, our results naturally translate to the other odd isotopologues of BaF. We also expect them to be relevant for species with similar structure, like isotopologues of CaF, SrF or radioactive RaF~\cite{GarciaRuiz2020} containing metal atoms with non-zero nuclear spin. Other examples where optical cycling and laser cooling of diatomic molecules with two nuclear spins have recently been considered include AlF~\cite{Hofsaess2021}, TlF~\cite{Grasdijk2020centrex} and YbOH~\cite{Pilgram2021}, but detailed studies of laser cooling forces have so far been lacking.

\begin{figure}[tb]
\includegraphics[width=.45\textwidth]{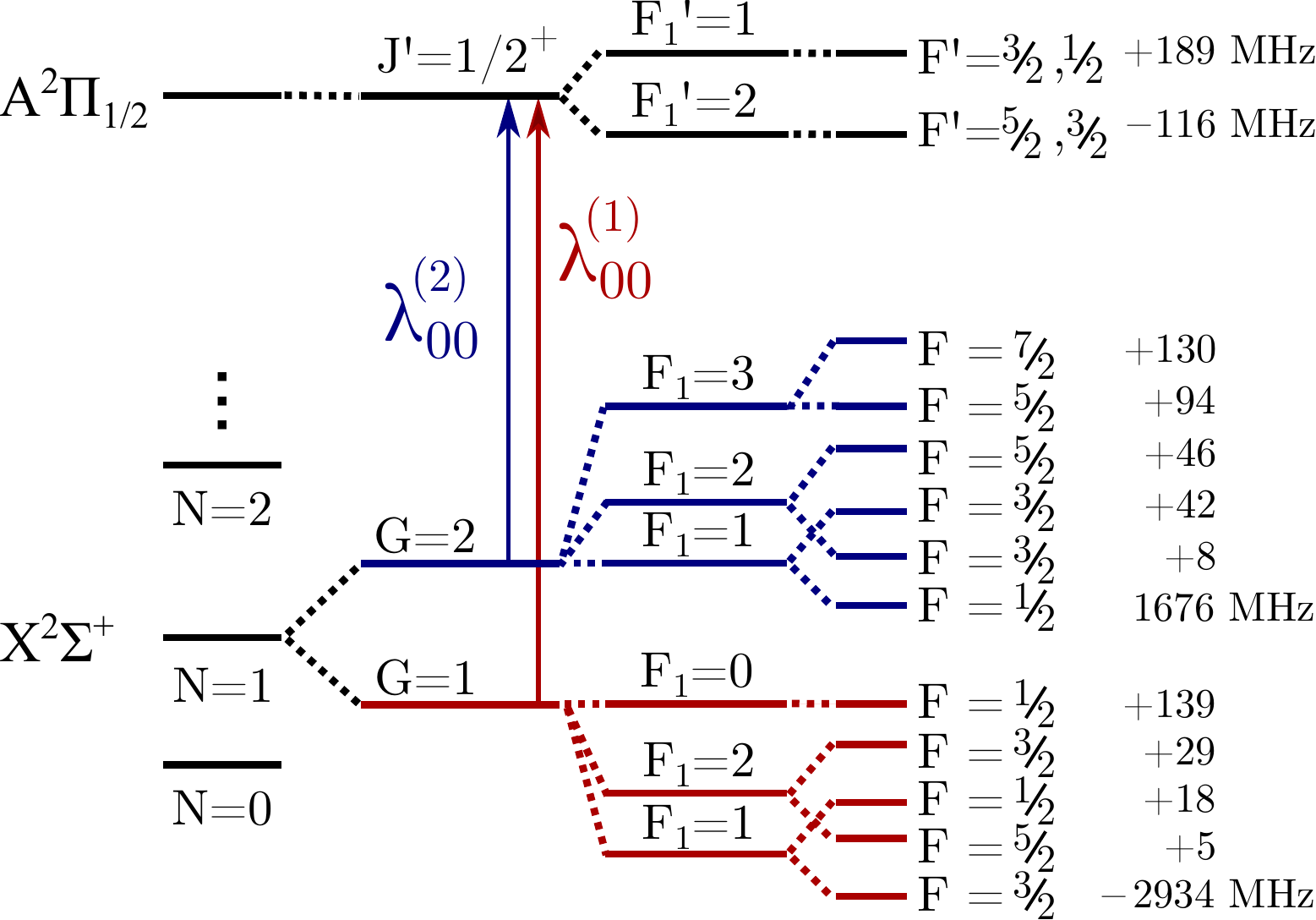} 
\caption{Relevant energy levels for the laser cooling of \baf\ including their relative energy shifts. As in other laser cooling schemes for molecules, rotational branching is suppressed using parity and angular momentum selection rules on the \gs, $N=1$ ground state to \exs, $J'^p=1/2^+$ excited state transition. Here, $N$ denotes the rotational quantum number of the ground state with parity $p=(-1)^N$. The total angular momentum of the excited state is given by $J'^p$, with $p=+$ denoting a state with positive parity. Angular momentum couplings involving the nuclear spins of the barium  ($I_{{}^{137}\mathrm{Ba}}=3/2$) and fluorine nucleus ($I_\mathrm{F}=1/2$) lead to a hyperfine structure described by the quantum numbers $G$, $F_1$ and $F$. For optical cycling and laser cooling, two lasers at wavelengths $\lambda_{00}^{(1)}$ and $\lambda_{00}^{(2)}$ around $860\,$nm, separated by $4.61\,$GHz, and including suitable frequency sidebands address the states of the $G=1,2$  ground state manifolds to realize a quasi-closed $64$ level system.}
\label{fig:spectrum1}
\end{figure}

\section{Level scheme}

\begin{figure}[tb]
\includegraphics[width=.41\textwidth]{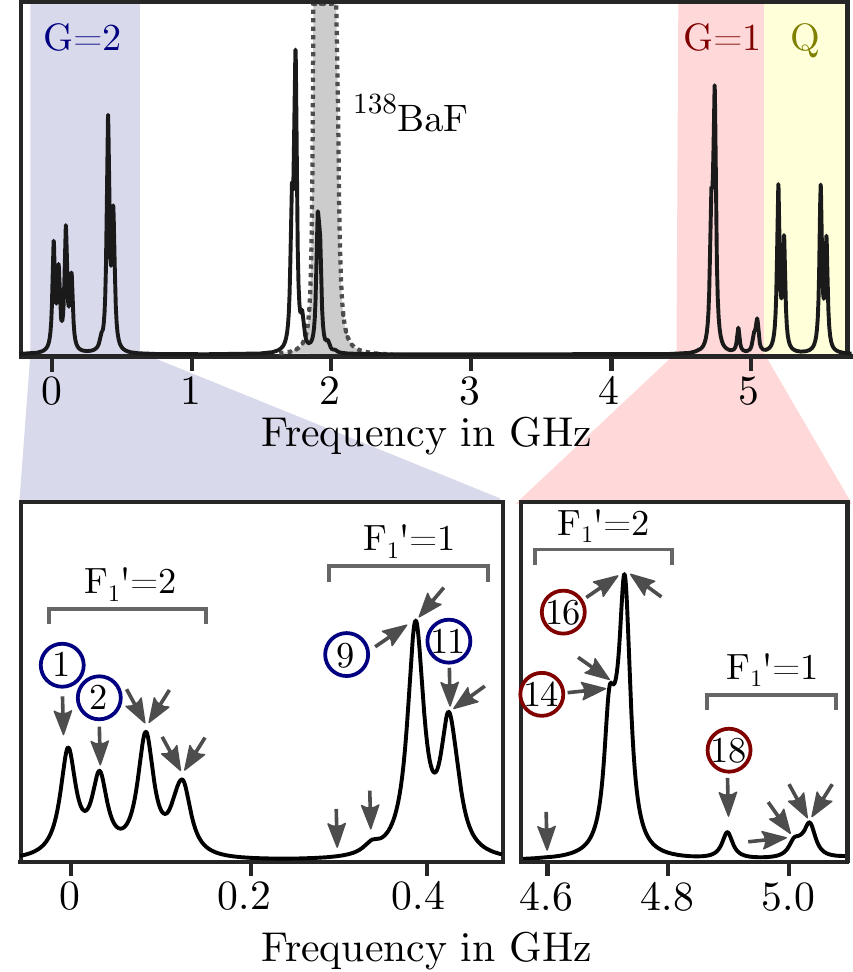} 
\caption{Spectrum of \baf\ in the region of the quasi-closed \gs, $N=1$ to \exs, $J'^p=1/2^+$ transition. Transition strengths include branching ratios and thermal occupation of the states~\cite{Steimle2011}. The blue and red shaded regions mark the transitions from ground states with $G=2$ and $G=1$, respectively. There are $22$ transitions in total, which are indicated by arrows in the respective insets. Horizontal bars label the corresponding excited states $F_1'=1,2$. In order to realize efficient cycling, it is sufficient to use seven frequency components to near-resonantly drive a subset of these transitions, which are labeled by their index in the plots. Further details on these transitions are given in Table~\ref{tab:transitions}. For reference, the corresponding transition in \baff\, is indicated by the dashed line. Other transitions shown belong to nearby branches. Of these, the yellow shaded Q branch is of particular concern as a loss channel, due to its proximity to the $G=1$ transitions.}
\label{fig:spectrum2}
\end{figure}

We start by recalling the relevant aspects of the structure of BaF. As ${}^{19}$F is the only stable fluorine isotope, there is a series of isotopologues of BaF, which are characterized by the various isotopes of barium. Of these, bosonic \baff\ is the most abundant one (71.7\%), followed by fermionic \baf\ (11.2\%). 

The former has been extensively discussed in the context of laser cooling, and optical cycling can be achieved by standard techniques~\cite{Fitch2021}. The \gs\ to \exs\ transition has a convenient wavelength of $\lambda=860\,$nm, where ample power is available from simple diode lasers. Moreover, the linewidth of the excited \exs\, state $\Gamma=2\pi\times2.84\,$MHz is sufficiently broad to realize fast photon cycling. The Franck-Condon factors in \baff\ are known to be highly diagonal, which strongly suppresses branching into unwanted vibrational states and thus limits the number of required repumping lasers~\cite{DiRosa2004,Chen2016,Chen2017,Albrecht2020}. As in other monofluorides, like SrF and CaF, rotational branching can be suppressed by exploiting parity and rotational angular momentum selection rules on the \gs, $N=1$ to \exs, $J'^p=1/2^+$ transition~\cite{Stuhl2008,Fitch2021}. Here, $p$, $N$ and  $J$ and denote the quantum numbers of parity, rotational and total angular momentum without nuclear spin, respectively

The further substructure of ground and excited states is determined by the fact that only the ${}^{19}$F atom carries nuclear spin $I_\mathrm{F}=1/2$, while the ${}^{138}$Ba atom has zero nuclear spin. This leads to a simple hyperfine structure with hyperfine levels $F=2,1,0,1$ in the $N=1$ ground state, and unresolved hyperfine levels $F'=0,1$ in the  $J'^p=1/2^+$ excited state~\cite{Albrecht2020}. The four individual transitions between these levels can be addressed by a single laser, with frequency sidebands created using a standard electro-optical modulator. This leads to a quasi-closed cycle that involves $4$ excited and $12$ ground states. This number of states includes magnetic sublevels, but neglects higher vibrational states. Although the number of states involved in the cycling decreases the scattering rate $\Rsc = \Gamma \times N_e/(N_e+N_g)=\Gamma /4$ and increases the saturation intensity $I_\mathrm{sat,eff} = 2 I_\mathrm{sat} N_g^2/(N_e+N_g) = 18\times I_\mathrm{sat}$~\cite{Fitch2021} in comparison to the values $\Rsc = \Gamma/2$ and  $I_\mathrm{sat}=0.584$ mW/cm$^2$ obtained for a two-level system with the same wavelength and linewidth, it is straightforward to realize optical cycling in \baff~\cite{Chen2017,Albrecht2020}. Based on this, several experiments aiming to realize laser cooling of \baff\ are currently underway~\cite{Chen2017,Aggarwal2018,Albrecht2020}. 

Here, we are interested in parity-violating effects caused by the weak nuclear interaction. In this context, it is important to observe that the single valence electron in \baff\ is strongly localized around the barium atom with $I_{{}^{138}\mathrm{Ba}}=0$. As a consequence, parity violation effects are negligible in this isotopologue~\cite{Altuntas2018}. In contrast to this, \baf\ features an additional nuclear spin $I_{{}^{137}\mathrm{Ba}}=3/2$, which is expected to result in a detectable parity violation signal~\cite{Demille2008}. In the following, we therefore discuss the consequences of this additional nuclear spin for optical cycling and laser cooling.

Because of the small mass difference, transition wavelengths, lifetimes and Franck-Condon factors for \baf, are nearly identical to the ones in \baff. Branching into higher vibrational states is thus again strongly suppressed~\cite{appendix}. Moreover, the \gs, $N=1$ to \exs, $J'^p=1/2^+$ transition still represents a rotationally-closed optical cycle. However, the additional nuclear spin of the barium atom leads to a significantly more complex hyperfine structure~\cite{Steimle2011}, which is summarized in Fig.~\ref{fig:spectrum1}.
 
While rotational levels in the \gs\ ground state can still be described by the rotational quantum number $N$ and parity $p=(-1)^N$, the electron spin $\Op{S}$ couples strongly to the barium nuclear spin to form an intermediate angular momentum $\Op{G}=\Op{S}+\Op{I}_{{}^{137}\mathrm{Ba}}$. This causes each rotational state to be split into two well-separated manifolds $G=1,2$. The intermediate angular momentum then couples to the rotational angular momentum to form $\Op{F}_1 = \Op{N} + \Op{G}$. This coupling scheme is known as Hund's case $(b_{\beta S})$. A second, weaker hyperfine interaction arises due to the fluorine nuclear spin $I_\mathrm{F}=1/2$, which leads to the total angular momentum $\Op{F}=\Op{F}_1+\Op{I}_F$ and many well-resolved hyperfine levels.  

For the excited state \exs, the total angular momentum without nuclear spin $\Op{J}$ known from regular Hund's case (a) is simply extended by adding both nuclear spins, i.e. $\Op{J} +\Op{I}_{{}^{137}\mathrm{Ba}}=\Op{F}_1$ and $\Op{F}_1 +\Op{I}_{F}=\Op{F}$. As in \baff\, the hyperfine interaction from $I_\mathrm{F}=1/2$ is too weak to be resolved in the excited state. 

The resulting spectrum for the \gs, $N=1$ to \exs, $J'^p=1/2^+$ cycling transition is shown in Fig.~\ref{fig:spectrum2}. The corresponding transitions are presented in Tab.~\ref{tab:transitions}. For further details on the calculation of their properties see Refs.~\cite{appendix, Steimle2011}. In total, ground and excited state hyperfine levels are connected by $22$ transitions, two of which are approximately dipole forbidden~\footnote{As the interaction strength between rotation $\Op{N}$ and intermediate angular momentum $\Op{G}$ is comparable to the hyperfine coupling of the second nuclear spin $\Op{I}_\mathrm{F}$, the effective Hamiltonian mixes different $F_1$ and cause a violation of the selection rule $\Delta F_1 = 0,\pm 1$, rendering two transitions only approximately forbidden.}.  

\section{Optical cycling}
Based on these considerations, we see that in order to realize optical cycling, two lasers are required to individually address the $G=1$ and $G=2$ ground states in \baf. The frequency difference between the two states is approximately $4.61\,$GHz, which can, for example, easily be covered using two offset-locked diode lasers. Similar pairs of lasers are required to address the repumping transitions for higher vibrational states.

Given these pairs of lasers, there are several possibilities to apply frequency sidebands that address the individual hyperfine levels. 

First, naturally, all $22$ transitions can be directly driven by individual laser sidebands. In this case, the optical cycle involves a total number of $N_g=48$  ground and $N_e=16$ excited states including magnetic sublevels. This yields a maximum scattering rate $\Rsc = \Gamma/4$ and a saturation intensity of $I_\mathrm{sat,eff} = 72\times I_\mathrm{sat}$. While $64$ states are involved in total, the expected maximum scattering rate thus remains unchanged from the situation in \baff\ and other monofluorides with a single nuclear spin. At the same time only a factor of four higher intensity than in \baff\ is required to saturate the cooling transition. This scaling is due to the large proportion of excited states and remains equally favorable when including additional vibrational states. 

However, driving all transitions requires very irregularly-spaced frequency spectra (see Fig.~\ref{fig:spectrum2}). In principle, such spectra could be generated using combinations of frequency shifting, spectral broadening, spectral engineering based on serrodynes~\cite{Rogers2011} or holographic algorithms~\cite{Holland2021}. However, bulk setups to achieve this are complex and a large fraction of laser power will inevitably be lost into off-resonant spectral components, or drive unwanted transitions nearby. Integrated broadband modulators could be used to generate more sophisticated waveforms, such as serrodynes, but in this case our simulations suggest significantly reduced scattering rates, due to the large number of states involved in the cycling~\cite{Holland2021}.

A straightforward way to reduce this large number of states would be to address only one of the two excited states $F_1'=1,2$ with  $N_e=6,10$, respectively. However, this does not significantly simplify the required frequency spectrum, while $\Rsc$ is even reduced and $I_\mathrm{sat,eff}$ increased.

For a practical scheme, it is therefore imperative to further reduce the number of required frequency components. In this context, the minimum requirement to achieve cycling is to ensure that all ground states are addressed to avoid the generation of dark states. 

\begin{table}[tb]
	\begin{tabular*}{0.45\textwidth}{l @{\extracolsep{\fill}} ccrc}
		\toprule
		index &  $\ket{G,F_1,F}$ & $\ket{F'_1,F'}$ & $f$ in MHz & $r$ \\
		\midrule
		\textbf{1} & $\ket{2,3,7/2}$ & $\ket{2,5/2},\ket{2,3/2}$ & $\mathbf{0.00}$ & $\mathbf{1.305}$\\ 
		\textbf{2} & $\ket{2,3,5/2}$  & $\ket{2,5/2},\ket{2,3/2}$ &  $\mathbf{35.32}$ & $\mathbf{0.975}$\\ 
		3 &   $\ket{2,2,5/2}$  & $\ket{2,5/2},\ket{2,3/2}$  & $83.65$ & 0.261\\ 
		4 &   $\ket{2,1,3/2}$  & $\ket{2,5/2},\ket{2,3/2}$  & $87.69$ & 1.335\\ 
		5 &   $\ket{2,2,3/2}$  & $\ket{2,5/2},\ket{2,3/2}$  & $121.28$ & 0.385\\ 
		6 &   $\ket{2,1,1/2}$  & $\ket{2,5/2},\ket{2,3/2}$  & $129.50$ & 0.724\\
		\addlinespace 
		7 &   $\ket{2,3,7/2}$ & $\ket{1,3/2},\ket{1,1/2}$  & $304.50$  & 0.000 \\ 
		8 & $\ket{2,3,5/2}$  & $\ket{1,3/2},\ket{1,1/2}$ &  $339.82$ & 0.119\\ 
		\textbf{9} &   $\ket{2,2,5/2}$  & $\ket{1,3/2},\ket{1,1/2}$  & $\mathbf{388.15}$ & $\mathbf{2.151}$\\ 
		10 &   $\ket{2,1,3/2}$  & $\ket{1,3/2},\ket{1,1/2}$ & $392.19$ & 0.914\\ 
		\textbf{11} &   $\ket{2,2,3/2}$  & $\ket{1,3/2},\ket{1,1/2}$  & $\mathbf{425.78}$ & $\mathbf{1.450}$\\ 
		12 &  $\ket{2,1,1/2}$  & $\ket{1,3/2},\ket{1,1/2}$  & $434.00$ & 0.424\\
		\addlinespace
		13 &  $\ket{1,0,1/2}$ & $\ket{2,5/2},\ket{2,3/2}$  & $4600.64$ & 0.004\\ 
		\textbf{14} &  $\ket{1,2,3/2}$  & $\ket{2,5/2},\ket{2,3/2}$  & $\mathbf{4710.46}$ & $\mathbf{1.453}$\\ 
		15 &   $\ket{1,1,1/2}$  & $\ket{2,5/2},\ket{2,3/2}$  & $4722.00$  & 0.417\\ 
		\textbf{16} &  $\ket{1,2,5/2}$  & $\ket{2,5/2},\ket{2,3/2}$  & $\mathbf{4734.88}$ & $\mathbf{2.389}$\\ 
		17 &   $\ket{1,1,3/2}$  & $\ket{2,5/2},\ket{2,3/2}$  & $4739.55$& 0.841\\
		\addlinespace 
		\textbf{18} &   $\ket{1,0,1/2}$ & $\ket{1,3/2},\ket{1,1/2}$  & $\mathbf{4905.14}$ & $\mathbf{0.309}$\\ 
		19 &   $\ket{1,2,3/2}$  & $\ket{1,3/2},\ket{1,1/2}$  &  $5014.96$& 0.171 \\ 
		20 &    $\ket{1,1,1/2}$  & $\ket{1,3/2},\ket{1,1/2}$ & $5026.50$ & 0.079 \\ 
		21 &   $\ket{1,2,5/2}$  & $\ket{1,3/2},\ket{1,1/2}$  & $5039.38$ & 0.257\\ 
		22 &   $\ket{1,1,3/2}$  & $\ket{1,3/2},\ket{1,1/2}$   & $5044.05$ & 0.143\\ 
		\bottomrule
	\end{tabular*}
	\caption{Relevant transitions between ground states $\ket{G,F_1,F}$ and excited states $\ket{F'_1,F'}$,  labeled with the same indices as in Fig. \ref{fig:spectrum2}. For laser cooling, transitions marked in bold are addressed by individual near-resonant frequency components. These frequency components also drive nearby transitions off-resonantly. For the derivation of the frequency shifts $f$ and relative strengths $r$ see Ref.~\cite{appendix}. }
	\label{tab:transitions}
\end{table}

As indicated in Fig.~\ref{fig:spectrum2} and Tab.~\ref{tab:transitions}, we thus choose to directly drive $7$ out of the total $22$ transitions that are, given their large relative strength, expected to account for most of the scattering. Transitions that are separated by less than $4\,\Gamma$ from the selected ones, are driven off-resonantly without adding extra frequency components. Apart from transition $18$, which must be addressed to avoid optical pumping into this state, other weaker transitions can be neglected. An inherent advantage of this choice is that possible losses into the Q branch are suppressed since the four transitions $19$-$22$ closest to this branch are not directly addressed (see Fig.~\ref{fig:spectrum2}). Most importantly, suitable laser sidebands to address this subset of transitions can be realized using standard bulk setups. 

\section{Doppler and sub-Doppler cooling}
Having established a quasi-closed level scheme, we now demonstrate how this scheme can be used to realize Doppler and sub-Doppler forces. To do so, we numerically solve the optical Bloch equations including all 64 levels, external magnetic fields and all frequency components discussed above~\cite{Devlin2018,Kogel2021code}. 

\subsection{Doppler cooling}
\begin{figure}[tb]
	\includegraphics[width=0.45\textwidth]{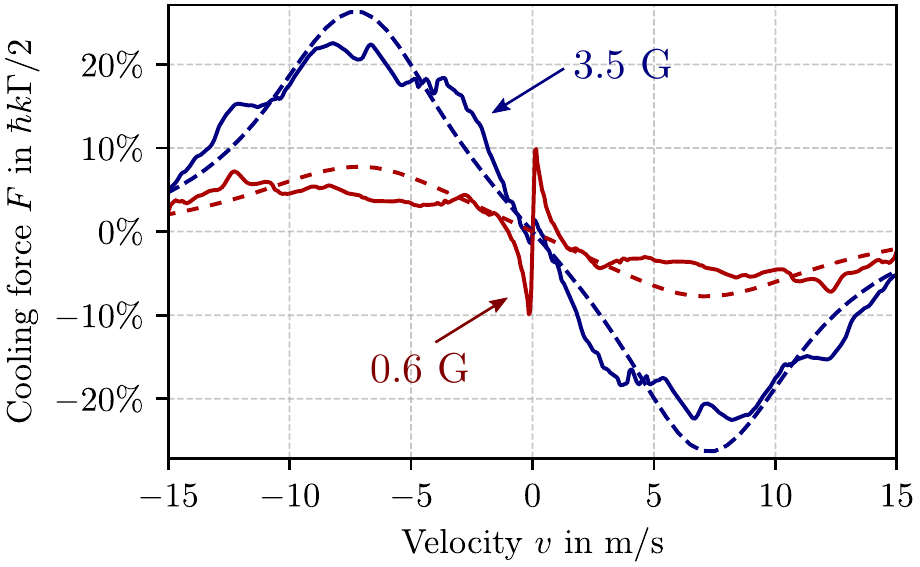} 
	\caption{Doppler cooling forces in a one-dimensional optical molasses configuration. These simulations are based on OBEs (solid lines) and rate equations (dashed lines) for high and low magnetic field strengths of $3.5\,$G (blue) and $0.6\,$G (red), respectively. The seven transitions highlighted in Fig. \ref{fig:spectrum2} are driven by laser sidebands, each with an intensity of $24.3\, $mW/cm$^2$ and red detuning of $\Delta = -2.5 \Gamma$. Note that this intensity is sufficient to saturate these transitions, and also to off-resonantly address other transitions nearby. The given intensity can easily be achieved using moderate laser powers, by using retro-reflected beams for cooling~\cite{Lim2018}. For small magnetic field strengths, sub-Doppler forces arise in the region of small velocities due to the magnetically-assisted Sisyphus effect.}
	\label{fig:doppler}
\end{figure}

The results for molecules exposed to a one-dimensional molasses configuration with two counterpropagating laser beams are shown in Fig.~\ref{fig:doppler}.

We destabilize dark magnetic sublevels using magnetic fields of $0.6\,$G or $3.5\,$G, respectively, which are applied under an angle of $60^\circ$ with respect to the linear polarization of the laser beams. As usual, the destabilization of the dark states can alternatively also be realized using sufficiently fast polarization switching. Each frequency component is red-detuned by $\Delta=-2.5\Gamma$ with an intensity of $24.3\, $mW/cm$^2$. A clear Doppler cooling force is visible, which reaches a maximum of $\sim 0.25\times \hbar k\Gamma/2$ for high magnetic fields. Here, $\hbar k\Gamma/2$ is the maximum force in a two-level system with the same wavelength and linewidth. 

Increasing the intensity does not necessarily increase the Doppler force further in type-II configurations with $F\ge F'$, such as the one considered here. Instead, as already observed in other species, the intensity has to be optimized specifically for each magnetic field strength, in order to match the given Zeeman shifts and Larmor frequencies of the states involved~\cite{BarryPhD,McNallyPhD}. Such optimizations lead to a large parameter space and are thus computationally costly using the OBEs. For faster optimizations of the parameters it is thus preferable to solve a simpler rate equation model~\cite{Fitch2021,Tarbutt2015}. As shown in Fig.~\ref{fig:doppler}, this model agrees well with the solution of the full OBEs but neglects all coherent effects. 

Note, that we do not expect the cooling force to reach the idealized limit $F=\hbar k \times \Rsc$ in a realistic system with many levels, since individual frequency components are inevitably both red- and blue-detuned to certain transitions at the same time, which is not taken into account in this simple estimate. In line with this, we observe that the maximum force is only slightly enhanced by about $20\,$\% when driving all transitions, rather than just the seven selected above. Significantly reducing the complexity of the setup by reducing the number of addressed transitions thus only leads to a minor decrease in the observed force. We anticipate that a similar approach will be applicable for many more molecular species that have so far seemed out of reach of laser cooling schemes.

\subsection{Sisyphus cooling}
In addition to Doppler cooling, it is well known that magnetically-assisted Sisyphus cooling techniques are very efficient in cooling molecules below the Doppler limit~\cite{Truppe2017}. These techniques have been shown to be a particularly powerful tool to collimate molecular beams for precision measurements of fundamental physics~\cite{Lim2018,Alauze2021}. As in atoms, these techniques rely on a position-dependent AC Stark shift and precisely timed, magnetically induced transitions between dark and bright states~\cite{Barry2014,Fitch2021}. In the following we study this behavior in \baf\ and realize strong cooling forces that require the scattering of a much lower number of photons than the Doppler forces discussed in the previous section.

A first sign of these Sisyphus forces is visible in the one-dimensional molasses configuration shown in Fig.~\ref{fig:doppler}. At small magnetic fields, slower remixing of dark states decreases the Doppler force. In contrast to this, the Sisyphus effect is enhanced in this regime since the rate of transitions between dark and bright states is better matched to the atomic motion through the standing wave intensity pattern formed by the cooling lasers. Notably, the direction of the Sisyphus force is opposite to the one of the Doppler force. This leads to a competition between Doppler cooling and Sisyphus heating, which is a general feature of red-detuned type-II systems~\cite{Barry2014}. 

\begin{figure}[tb]
	\includegraphics[width=0.45\textwidth]{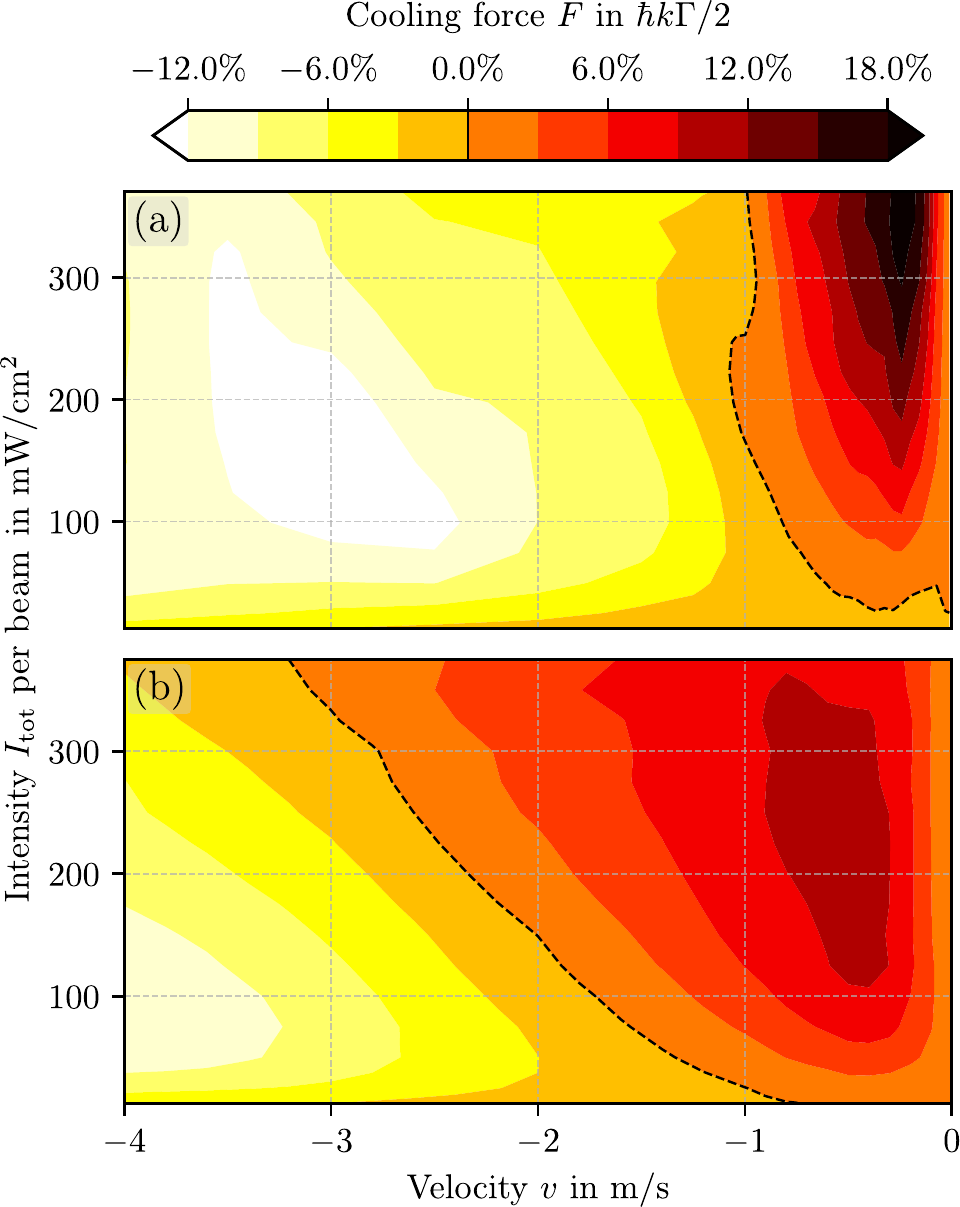} 
	\caption{Sub-Doppler cooling forces in one- (a) and two-dimensional (b) optical molasses as a function of the total intensity per beam $I_\mathrm{tot}$. The seven individual frequency components are blue-detuned by $\Delta = 2.5 \Gamma,$ and the magnetic field strength is $1.2\,$G. In two dimensions, the projection of the forces onto different velocity directions are approximately the same and we average them to yield the forces shown. While the two-dimensional Sisyphus forces are stronger at low intensity and reach their maximum value for $I_\mathrm{tot}\approx 200\,$mW/cm$^2$, the one-dimensional forces continue to increase for all intensities studied, and reach $0.26\times \hbar k\Gamma/2$ for $I_\mathrm{tot}=650\,$mW/cm$^2$. Notably, the range of the force significantly increases in a two-dimensional configuration.}
	\label{fig:subdoppler}
\end{figure}

To turn this heating effect into cooling and further maximize the Sisyphus forces, we detune the lasers $\Delta=2.5\Gamma$ to the blue and apply a low magnetic field of $B=1.2\,$G. A systematic study of the resulting forces as a function of velocity $v$ and total intensity $I_\mathrm{tot}$ per beam is shown in Fig. \ref{fig:subdoppler}a, where a maximum force of $0.26\times \hbar k\Gamma/2$ is reached for $I_\mathrm{tot}=650\,$mW/cm$^2$ at a velocity of $v=0.2\,$m/s. 

Interestingly, we find that these sub-Doppler forces are enhanced by up to $33\,$\% in our reduced level scheme. Driving only $7$ out of $22$ transitions in total can thus, counter-intuitively, have a positive effect on the actually observed forces, as undesired coherent resonances between individual excitation pathways are reduced.

We have also studied the extension of this scenario to two dimensions. For this, the same counterpropagating laser beams are also applied in the second transversal direction, with their linear polarizations orthogonal to the ones of the original beams. The magnetic field strength is aligned under an angle of $45^\circ$ with respect to the laser polarizations and remains of equal strength as before. Interference between the beams results in a two-dimensional standing wave with periodic gradients in both intensity and polarization. Therefore, the magnetically-induced force is combined with a polarization gradient cooling mechanism, where non-adiabatic transitions between bright and dark states are induced by the molecule's motion through the changing polarization. The resulting force is found by averaging over many relative laser phases corresponding to different initial positions of the molecule within a square with a sidelength of one laser wavelength~\cite{Devlin2018}. Note that the projections of this force onto the different velocity directions are approximately the same, which is why we only discuss its absolute value in the following.

While the two-dimensional Sisyphus forces are slightly stronger at low intensity and reach a maximum at $I_\mathrm{tot}\approx 200\,$mW/cm$^2$ per beam, the one-dimensional forces continue to increase with intensity over the whole range of intensities studied. We find that the velocity range over which the force is active, and thus the phase space acceptance of the overall cooling method, increase in the two-dimensional configuration. The same behavior has previously also been observed for YbF molecules~\cite{Alauze2021}.

Notably, as the molecules spent a large amount of time in dark states, we observe a reduction of the excited state population $n_e$, which becomes significantly lower than in Doppler cooling. As a consequence, repumping only the $\nu=1$ vibrational level will be sufficient for the sub-Doppler cooling of the molecules. Our simulations indicate that if such repumping is performed with an intensity equal or exceeding that of the main cooling laser, the magnitude of the sub-Doppler forces remains nearly unchanged.

In particular, for small velocities $v\approx0$, $n_e$ can further drop to only $\sim2.5\,\%$. The coldest molecules thus exhibit a very small heating rate due to the low photon scattering rate $R_\mathrm{sc}=\Gamma n_e$. A low equilibrium temperature can thus be achieved in a way that is reminiscent of velocity-selective coherent population trapping~\cite{Alauze2021}. 

\section{Coherent bichromatic forces}
Significantly larger forces than spontaneous Doppler and sub-Doppler forces can be realized using stimulated techniques. In particular, bichromatic forces (BCF) have been successfully implemented using both atoms~\cite{Soeding1997,Chieda2012} and molecules~\cite{KozyryevBichromatic2018,GalicaDeflection2018}. 

These forces can be realized using counterpropagating beat note trains, which are formed using two laser beams. Each beam contains two frequencies $\omega\pm\delta$ that are symmetrically detuned around the resonance frequency $\omega$. In an example two-level system, the relative phase and intensity of the two beams can be adjusted to realize a situation, where each beat note corresponds to a $\pi$-pulse. This leads to cycles of absorption from one beam and stimulated emission into the other beam. In an ideal situation, it is thus possible to transfer a momentum of $2\hbar k$ per optical cycle at a rate $\delta/2\pi$ that exceeds $\Gamma$, leading to forces that can be significantly larger than spontaneous forces. The maximum force $F_\mathrm{BCF}=\hbar k \delta/\pi$ is achieved when the Rabi frequency is given by $\Omega_\mathrm{R}=\sqrt{3/2}\, \delta$ and the relative phase of the counterpropagating beams is adjusted to $45^\circ$. As $\delta$ can be very large, there is no immediate fundamental limit to this force. However, a practical limit arises from the laser intensity $\sim \delta^2$ that is required to reach the ideal $\Omega_\mathrm{R}$. 

For molecules, bichromatic forces are expected to be a particularly powerful tool, as branching into unwanted states is strongly suppressed by the stimulated relaxation. Moreover, there is a reduced overall scattering rate and multiple coherent momentum transfer cycles occur per spontaneous scattering event.

However, for molecules with many internal levels the $\pi$-pulse condition is challenging to fulfill for all involved transitions at once. Thus, a careful choice of intensity is required to maximize the force. The bichromatic detuning $\delta$ also must be large enough to cover all levels involved. In  \baf\ this can not be easily achieved as the $G=1,2$ ground state manifolds are separated by several thousand linewidths $\Gamma$. A large enough detuning to cover this energy splitting would thus require unrealistically high laser powers. In addition, it would also lead to undesired losses, as other unwanted transitions nearby are addressed off-resonantly. Even if two lasers are again used to individually address $G=1,2$ as in Doppler and sub-Doppler cooling, a similar problem arises in connection with the two excited state manifolds $F'_1=1,2$. Here, the required detuning to cover their energy splitting of over $300\,$MHz (see Fig. \ref{fig:spectrum1}) would again mandate prohibitively large laser powers. 

\begin{figure}[tb]
	\includegraphics[width=0.45\textwidth]{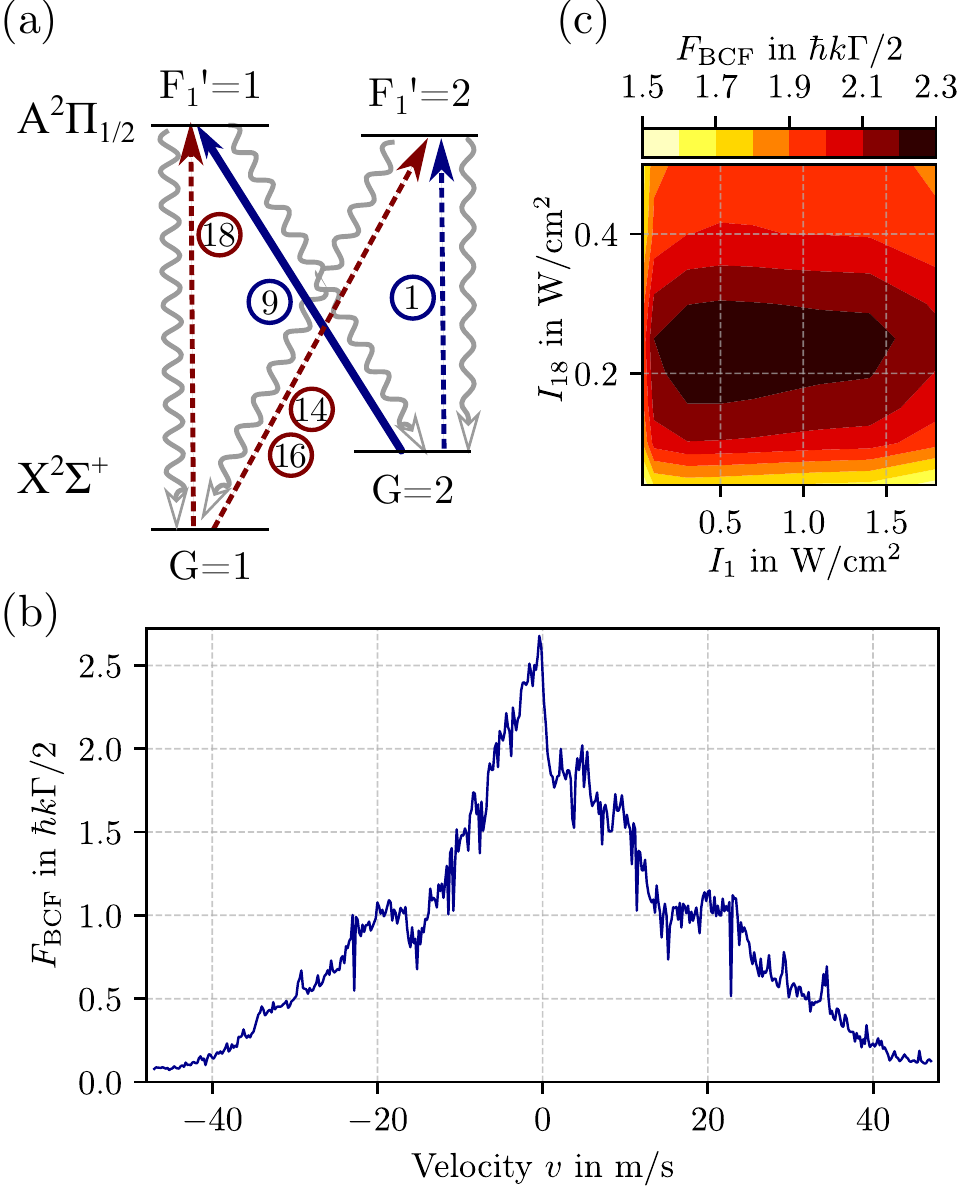} 
	\caption{Realizing bichromatic forces. (a) Simplified level scheme including the strong bichromatic laser field (centered on transition $9$ and addressing transitions $8-12$) and direct (transitions $1$ and $18$) and indirect repumpers (transitions $14$ and $16$). (b) Resulting bichromatic force profile. (c) Optimization of the peak force as a function of the intensities $I_{1}$ and $I_{18}$ of the direct repumpers on transitions $1$ and $18$, respectively. As transition $1$ and the BCF light field address different sublevels of $G=2$ due to selection rules, the force magnitude depends only weakly on $I_1$. In contrast to this, transition $18$ directly addresses one of the excited states of the coherent bichromatic cycle, leading to a stronger dependence of the force on $I_{18}$.}
	\label{fig:bichromatic}
\end{figure}

As it is not possible to easily cover all ground and excited states, we again follow the strategy to address only a subset of the transitions. To illustrate this, in Fig.~\ref{fig:bichromatic}a we show a reduced level scheme involving the $G=1,2$ and $F'_1=1,2$ manifolds~\footnote{We note that, while the simplified level scheme shown in Fig.~\ref{fig:bichromatic}a is similar to a scheme that was recently proposed for the realization 
of large molasses-like bichromatic cooling forces~\cite{Wenz2020a}, such forces can not be realized in \baf\ efficiently due to the various loss channels discussed in the main text.}. In principle, all four possible transitions in this reduced level scheme could be suitable for applying the BCF. Simulating the bichromatic forces for each subsystem isolated from the others, we find that driving the transitions connecting $G=2$ and $F'_1=1$ exhibit the largest forces. In contrast to this, addressing $G=1$ with the high-intensity bichromatic field is disadvantageous as molecules can be lost through the nearby Q-branch (see Fig.~\ref{fig:spectrum2}). 

In order to achieve the most efficient bichromatic forces, we center the lasers around the frequency of transition $9$ and optimize the detuning of the bichromatic laser sidebands, which must be carefully chosen.
On the one hand, a large detuning is desired to achieve large forces and address all transitions from $G=2$ to $F_1'=1$. On the other hand, if this detuning becomes comparable to the splitting of the excited states $F'_1=1,2$, the force vanishes~\cite{Aldridge2016}. Previous estimates for such a situation based on a simplified lambda system, consisting of two ground and one excited state, suggest that in our case the ideal detuning is around $\delta=28\,\Gamma$. Neglecting the weak transition $8$, this value is also in the large-detuning regime with respect to the $G=2$ hyperfine splitting of $16\,\Gamma$, and is thus well-suited to realize BCFs. 

However, applying the BCF only to a subset of states can lead to losses into dark states that need to be addressed. We find that only a small fraction of the population temporarily leaves the bichromatic cycling scheme via $F'_1=1$ to $G=1$. This population can easily be returned to the initial $G=2$ state using low-power repumpers on the strong transitions $14$ and $16$ and subsequent spontaneous decay from the other excited state $F'_1=2$ back into $G=2$. This indirect repumping is especially efficient as the bichromatic cycling is not disturbed by the incoherent spontaneous decay. Additional losses are possible due to optical pumping into the two ground states belonging to the dipole-forbidden transitions $7$ and $13$ (see Fig. \ref{fig:spectrum2}). Consequently, the population must be directly pumped out of these states back into the bichromatic cycle. To achieve this, we add weak direct repumpers on the two allowed transitions $1$ and $18$. As summarized in Fig. \ref{fig:bichromatic}a, the resulting scheme is very similar to the scheme discussed for the spontaneous forces.

The resulting force profile is shown in Fig. \ref{fig:bichromatic}b. It is calculated using the OBEs for the full $64$ level system, including the bichromatic laser components $\omega\pm\delta$ and the indirect and direct repumping sidebands tuned to their respective resonances. To remix dark magnetic sublevels, a magnetic field with strength $B=20\,$G and angle of $60^\circ$ to the linear laser polarizations is applied. 

To optimize the force, the intensity per bichromatic laser component is first estimated as $I=3I_\mathrm{sat}\delta^2/(\kappa\Gamma)^2$ to match the $\pi$-pulse condition using the quadrature-summed Rabi frequency. This involves the quadrature sum of the matrix elements of the electric dipole operator which gives $\kappa=1/\sqrt{3}$ for each single excited sublevel and is the same as for \baff, CaF and SrOH~\cite{Aldridge2016}. Numerically, we find an optimal intensity of $I=4.74\,$W/cm$^2$, which is $14\,$\% larger than this estimated value. Note that, in comparison to CaF, the required intensities for the same detuning $\delta/\Gamma$ are an order of magnitude smaller for BaF, leading to significantly less stringent requirements in terms of laser powers. The intensities of the direct repumping sidebands are set to $1.5\,$W/cm$^2$, while the ones of the indirect pumping sidebands addressing the transitions $1$ and $18$ are carefully adjusted to $I_1=0.25\,$W/cm$^2$ and $I_{18}=0.5\,$W/cm$^2$, respectively (see Fig. \ref{fig:bichromatic}c)~\footnote{Note that while the BCF light field addresses all ground states in $G=2$, transition 7, which shares its ground state with the direct repumper on transition 1, is dipole forbidden. This avoids any interference between the direct repumper and the BCF light field.}. 
 
While the velocity range covered by the BCF is smaller than previously observed in simpler molecular systems~\cite{Aldridge2016}, it exceeds Doppler and Sisyphus forces by more than one order of magnitude.

\begin{figure*}[tb]
\centering
\includegraphics[width=0.90\textwidth]{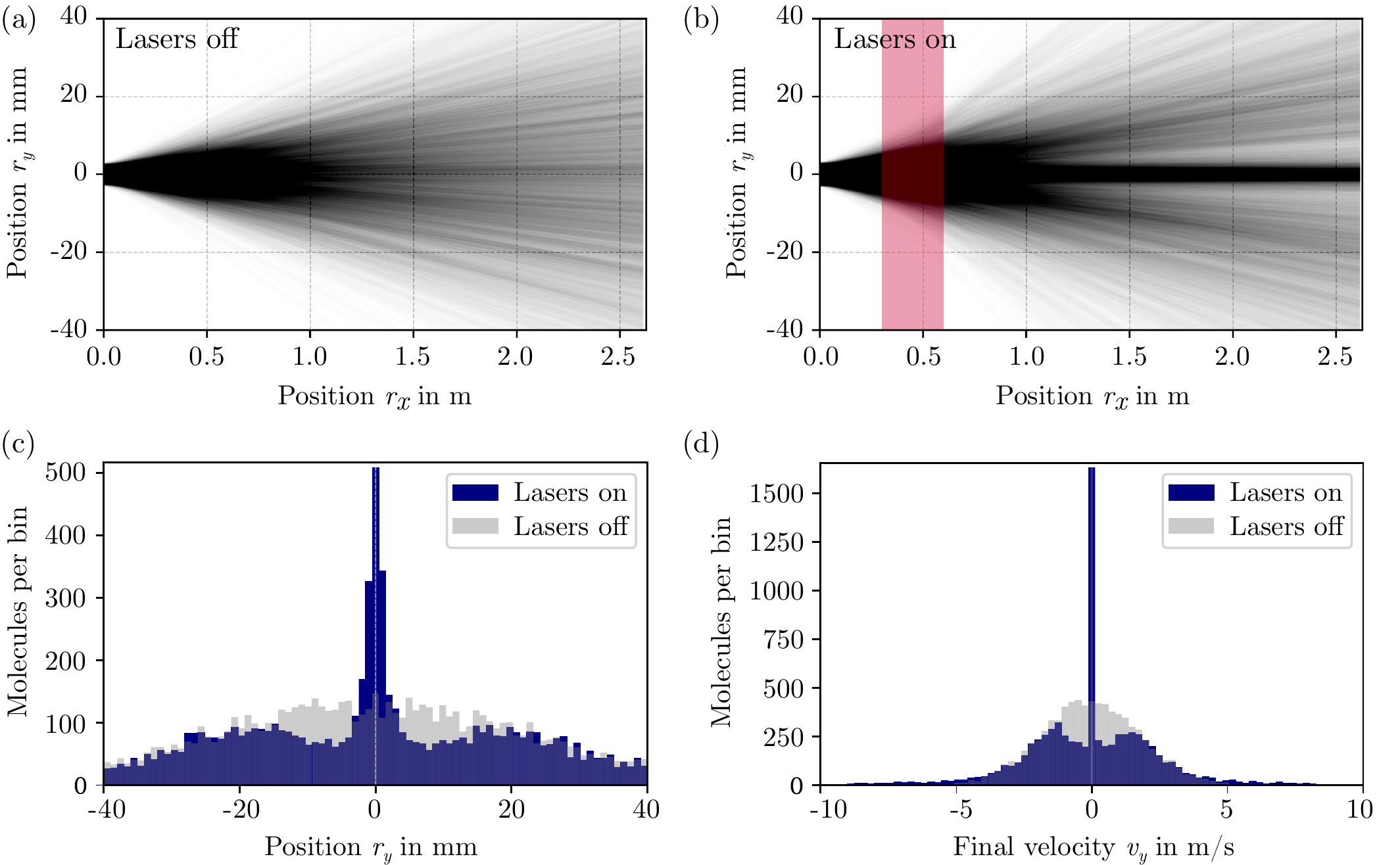} \\
\caption{Molecular trajectory simulations without cooling (a) and with two-dimensional transversal laser cooling (b). The red shaded region in (b) indicates the $30\,$cm-long laser cooling region. Following the cooling region, the molecules propagate freely for another two meters, during which the actual parity violation measurements can be performed~\cite{Altuntas2018}. (c) Position and (d) velocity distribution following this free propagation. Molecules within the range of the Sisyphus cooling forces accumulate at ultracold temperatures, leading to a focusing of the beam.}
\label{fig:exampleapplication}
\end{figure*}

\section{Application to tests of fundamental symmetries}
In the following, we illustrate the prospects for the cooling schemes to enhance precision experiments. 

There has recently been great interest in using \baf\ and similar molecules as probes for nuclear-spin-dependent parity violation~\cite{Demille2008,Norrgard2019}. By tuning rotational states of opposite parity into resonance using moderate magnetic fields, it is possible to enhance parity-violating effects by many orders of magnitude to provide a detailed window into the physics of the weak interaction and search for physics beyond the Standard Model~\cite{Demille2008,Dzuba2017}.

In a recent experiment the potential of this approach has been demonstrated using a beam of \baff\ molecules~\citep{Altuntas2018}. However, as discussed before, due to the vanishing barium nuclear spin, the parity-violating effects in this isotopologue are effectively zero. For the next generation of experiments, it will thus be necessary to implement such measurements also with \baf, where the finite nuclear spin $I_{{}^{137}\mathrm{Ba}}=3/2$ leads to significant parity-violating effects. Due to the low abundance of \baf, a key challenge in realizing such experiments is the production of a sufficiently intense molecular beam. In addition, this beam has to remain collimated while it propagates through a a several meter long apparatus. 

As a simple figure of merit for our cooling schemes, we thus characterize the gain of \baf\ molecules in the $N=1$ state for such a beam experiment. All other required experimental procedures and state preparation strategies have been described in detail elsewhere~\cite{Altuntas2018,AltuntasPRA}. 

As a starting point for our example, we use a buffer gas beam source. Our current source is based on a $4\,$K helium dewar cryostat and produces a flux of more than $10^9$ \baf\ molecules per steradian per pulse in the \gs, $N=1$ state~\cite{Albrecht2020}. The resulting BaF molecular beam has a forward velocity of $190\,$m/s, a longitudinal temperature of $10\,$K and, following collimation, a transversal temperature of approximately $100\,$mK. This corresponds to about two orders of magnitude more flux and a factor of three lower velocity as achieved using supersonic beam sources for BaF~\cite{Altuntas2018,RahmlowPhD}. Note that this forward velocity could be further lowered and the molecular flux increased using recently demonstrated quasi-continous buffer gas sources~\cite{Shaw2020}, as well as using the latest generation of closed-cycle cryocoolers reaching operational temperatures of below $2\,$K~\cite{Augenbraun2020,Jadbabaie2020}.

In a first step, we model two-dimensional transversal Sisyphus cooling of this molecular beam. For this, we study a $30\,$cm long cooling region that is located $30\,$cm behind the last collimating aperture of the molecular beam source. This interaction region is created using three elliptic laser beams that are passed through a glass cell vacuum chamber in parallel, and retro-reflected to form the vertical and horizontal transversal cooling beams. Their Gaussian $1/e^2$ intensity radii are $5\,$cm along the molecular beam axis and $1.5\,$mm in the perpendicular direction. The total laser power is $1\,$W, which is equally distributed among the three beams and the seven individual transitions, yielding $47.6\,$mW per frequency component in each beam.

Our configuration results in a maximum acceleration of around $2.7\,$km/$\mathrm{s}^2$, which is on the same order of magnitude as seen in simulations for YbF~\cite{Fitch2020methods}. We note that, in principle, similar results can also be obtained with even lower laser powers, by passing smaller beams through the cell more often~\cite{Alauze2021}. 

In Fig.~\ref{fig:exampleapplication}a and b, we show example molecular trajectory simulations, where the transversal laser cooling is shown to produce a highly-collimated molecular beam. Notably, due to the efficiency of sub-Doppler cooling, the molecules only scatter around $1000$ photons in the cooling region, such that only the $\nu=1$ vibrational level needs to be repumped. 

To further characterize the number of molecules that have been cooled to ultracold transversal temperatures, we plot in Fig.~\ref{fig:exampleapplication}c and d the position and velocity distributions after propagation for both cooled and uncooled beams. We find that even after $2\,$m of free propagation, around $16\%$ of the molecules can be found within a radius of $5\,$mm around the molecular beam axis and a transversal velocity smaller than $1\,$cm/s. This number of ultracold molecules corresponds to an enhancement of $10^4$ compared to the uncooled beam. This focusing of the beam could be made even more efficient by combining laser cooling with a magnetic lens~\cite{Alauze2021}. 

The spontaneous forces discussed so far are powerful tools to collimate a molecular beam. However, as mass and wavelength of BaF are comparably high, it will be challenging to use them to slow down a molecular beam. Given the recoil velocity of  $2.94\,$mm/s  for the \gs\ to \exs\ transition, such slowing would require the scattering of approximately $65000$ photons, the repumping of several vibrational states~\footnote{In this case, a minimum of $48$ more ground states need to be taken into account leading to a reduced scattering rate $R_\mathrm{sc}=\Gamma/7$} and a distance of at least $3\,$m. Moreover, similar to yttrium monoxide (YO)~\cite{Yeo2015}, BaF exhibits a potential loss channel through an intermediate $\Delta$ state that lies energetically lower than the \exs\ state~\cite{Yeo2015}. While the precise value of the branching ratio governing this loss is still an open question~\cite{Hao2019,Albrecht2020}, it is likely going to become relevant if more than $10^4$ photons are scattered.

More efficient slowing can be realized using the bichromatic techniques discussed above. Due to the small diameter of the molecular beam following transversal laser cooling, these forces can be applied with realistically achievable laser powers. In order to bring the molecules from $190\,$m/s to rest, a frequency-chirped scheme must be introduced to shift the force along the velocity axis as the velocity of the molecules is reduced. This can be achieved by adding a frequency offset $\pm \Delta \omega$ for the co- and counterpropagating beams, respectively~\cite{Chieda2012}. Assuming a mean force of $2.2\times \hbar k \Gamma/2$, slowing to rest is achievable within a slowing time of $t=3.27\,$ms and distance of $d= 31\,$cm, which is about $10$ times shorter than for radiation pressure frequency-chirped slowing. Given the Franck-Condon factors in BaF, we expect this force to be influenced by losses to higher vibrational states $\nu\ge1$. Although the respective mean scattering rate is reduced to $R_\mathrm{sc}=0.16\,\Gamma$, the characteristic times until the molecule is lost to $\nu=1$ or $\nu=2$ are $t_\mathrm{loss}=11\,\mu$s and $0.4\,$ms, respectively. This necessitates the use of two additional pairs of repumping lasers. 

Our simulations suggest that the maximum force in Fig.~\ref{fig:bichromatic}b remains equally favorable if the intensity deviates by not more than $15\,$\% from its optimum value. In the Gaussian intensity distribution of a laser beam with a full width at half maximum $\mathrm{FWHM}$ this situation is given for a circle with radius $r\le \mathrm{FWHM}/4$. Assuming a total laser power of $1\,$W ($2\,$W) the required intensity for the bichromatic scheme can be achieved for $\mathrm{FWHM}=2\,$mm ($2.8\,$mm). With this, we conservatively estimate that about $5\,$\% ($11\,$\%) of the ultracold molecules produced using transversal cooling can be efficiently slowed, yielding approximately $10^4-10^5$ molecules near standstill.  

We thus conclude that both the transversally collimated molecular beam with high flux and the slowed sample of ultracold molecules are both well-suited to investigate parity violation in \baf\ in the future, with the collimated beam being particularly straightforward to implement in current experiments. 

\section{Conclusion}
We have demonstrated a scheme for the optical cycling and laser cooling of \baf\ molecules. This scheme constitutes an important new tool for studies of fundamental physics using this and other molecular species. In particular, our scheme uses realistically achievable laser configurations and powers, and can readily be implemented in existing experiments~\cite{Albrecht2020,Altuntas2018}. In addition to a more intense molecular beam, our scheme could also facilitate more efficient preparation of the internal molecular states, as well as their rapid detection with near unit efficiency. Going beyond precision measurements, the ability to cool both \baf\ and the simpler \baff\ could facilitate studies of quantum statistical effects on chemical reaction rates and of state-to-state dynamics based on isotope-exchange interactions~\cite{Balakrishnan2016,Tomza2015}. On the theory side, the many states involved may offer many opportunities to engineer molasses techniques based on specific dark states to realize even larger cooling forces.
It will be interesting to explore how far the cooling techniques presented here can be pushed to make trapping of \baf\ molecules a viable option in the future.

\begin{acknowledgements}
We are indebted to Tilman Pfau for generous support and thank Timothy Steimle for sharing his computer code for the spectral analysis of \baf~\cite{Steimle2011}. This project has received funding from the European Research Council (ERC) under the European Union’s Horizon 2020 research and innovation programme (Grant agreement No. 949431), the Vector Stiftung, the RiSC programme of Ministry of Science, Research and Arts of the State of Baden-W\"urttemberg, and the Carl Zeiss Foundation.
\end{acknowledgements}

\appendix
\section*{Appendix}
\label{appendix}

\section{Franck-Condon factors}
A procedure to estimate the Franck-Condon factors based on Morse potentials has been outlined in our previous work on \baff~\citep{Albrecht2020}. To find the corresponding values for \baf\, we employ the usual mass scaling relations for the molecular constants, which are in good agreement with direct spectroscopic data~\cite{Steimle2011}. The results are summarized in Table~\ref{tab:FCF}.\\

\begin{table}[htb]
\begin{tabular*}{0.45\textwidth}{c @{\extracolsep{\fill}} cccc}
\hline 
 & $A(\nu'=0)$ & $A(\nu'=1)$ & $A(\nu'=2)$ & $A(\nu'=3)$ \\ 
\hline 
$X(\nu=0)$ & $0.9738$ & $0.0261$ & $9\times 10^{-5}$ & $7.2\times10^{-6}$ \\ 
$X(\nu=1)$ & $0.0254$ & $0.9206$ & $0.0537$ & $3.6\times10^{-4}$ \\ 
$X(\nu=2)$ & $8.2\times 10^{-4}$ & $0.0508$ & $0.8649$ & $0.0825$ \\ 
$X(\nu=3)$ & $7.1\times 10^{-6}$ & $0.0025$ & $0.0760$ & $0.8074$ \\  
\hline 
\end{tabular*} 
\caption{Franck-Condon factors of the $A^2\Pi_{1/2}(\nu')\rightarrow X^2\Sigma(\nu)$ transition in \baf.}
\label{tab:FCF}
\end{table}

\section{Effective g-factors}
In order to model Larmor precession and mixing between adjacent ground-state magnetic sublevels in the OBEs, the Zeeman shifts of the \baf\ energy levels must be evaluated. We consider the two major contributions to the Zeeman Hamiltonian~\cite{Brown2003}, which involve the electron spin g-factor $g_S=2.0023$ and the orbital g-factor $g'_L=1$. The resulting energy shifts exhibit an approximately linear behavior for small magnetic field strengths such as the ones considered here. The respective slopes of the eigenenergies $\Delta E_\text{Z} = g_\text{eff}\, m_F \mu_B B$ define the effective g-factors $g_\text{eff}$ that are used in the OBEs. They are summarized in Table~\ref{tab:g_eff} for reference. Here,  $\mu_B=\frac{e\hbar}{2m_e}$ is the Bohr-magneton with the electron mass $m_e$.

\begin{table}[htb]
	\centering
	\begin{tabular*}{0.17\textwidth}{l@{\extracolsep{\fill}} r}
		\toprule
		 $\ket{G,F_1,F}$ & $g_\text{eff}$\\
		\midrule
		 $\ket{1,1,3/2}$ & -0.251\\
		 $\ket{1,2,5/2}$ & -0.205\\
		$\ket{1,1,1/2}$ & -0.439\\
		 $\ket{1,2,5/2}$ & -0.241\\
		 $\ket{1,0,1/2}$ & 0.063\\
		 \addlinespace
		 $\ket{2,1,1/2}$ & 1.043\\		
		 $\ket{2,2,3/2}$ & 0.447\\
		 $\ket{2,1,3/2}$ & 0.578\\
		 $\ket{2,2,5/2}$ & 0.325\\
	 $\ket{2,3,5/2}$ & 0.395\\
		 $\ket{2,3,7/2}$ & 0.286\\
		\bottomrule
	\end{tabular*}
		\caption{Effective g-factors $g_\text{eff}$ for the eigenstates of the ground state \gs$(N=1$) for \baf.}
	\label{tab:g_eff}
\end{table}

\section{Transitions and branching ratios}
The ground state \gs, $N=1$ and the excited state \exs, $J'=1/2^+$ in \baf\ contain $11$ and $4$ hyperfine levels, respectively. As the frequency splitting of the excited states $\ket{F'_1,F'=F'_1\pm 1/2}$ for $F'_1=1,2$ is not resolved~\cite{Steimle2011}, this leads to a total of $22$ transitions. We list the properties of these transitions in Tab.~\ref{tab:transitions}. To calculate their frequency and strength, we express all states in the Hund's case (a) basis, set up the effective Hamiltonian, and calculate the matrix elements of the electric dipole moment operator $\Op{d}$ by using the Wigner-Eckart theorem~\cite{Brown2003},
\begin{multline}
	\braket{\eta',F',m'_F | \mathrm{T}^1_p(\Op{d}) | \eta,F,m_F} 
	= (-1)^{F'-m'_F} \sqrt{2F'+1} \\
	\times \Wj{F'}{1}{F}{-m'_F}{p}{m_F} 
	\braket{\eta',F' || \mathrm{T}^1_p(\Op{d}) || \eta,F} ,
\end{multline}
for the eigenstates $\ket{\eta',F',m'_F}$ and $ \ket{\eta,F,m_F}$. Here, $\eta$ and $\eta'$ represents the residual angular momentum quantum numbers of the ground and excited state, respectively, $\Op{d}$ is represented as a spherical tensor $\mathrm{T}^1_p(\Op{d})$, and $p$ denotes the laser polarization. The squares of the reduced matrix elements $\braket{\eta',F' || \mathrm{T}^1_p(\Op{d}) || \eta,F}$ are listed in Table~\ref{tab:branchingratios} and correspond to the probability of a single hyperfine excited state to decay to one of the ground hyperfine states. Consequently, each column is normalized to one. The selection rules are $\Delta F=0,\pm1$ and $\Delta F_1=0,\pm1$, where the latter is only approximately fulfilled, since eigenstates mix different $F_1$.

The relative transition strength $r$ in Table~\ref{tab:transitions} determines the amplitude of a single transition line in the spectrum in Fig.~\ref{fig:spectrum2} and is calculated by summing up all matrix elements $|\!\braket{\eta',F',m'_F | \mathrm{T}^1_p(\Op{d})| \eta,F,m_F}\!|^2$ of all sublevels belonging to the transition between the respective hyperfine excited $F'$ and ground state $F$. As this involves multiple magnetic sublevels $m'_F$, the relative transition strengths $r$ are weighted with the factor $(2F'+1)$ with respect to the reduced branching ratios in Table \ref{tab:branchingratios}, and can therefore be calculated by
\begin{align}
	r = (2F'+1)\times  |\!\braket{\eta',F' || \mathrm{T}^1_p(\Op{d}) || \eta,F}\!|^2 .
\end{align}

\begin{table}[htb]
	\begin{tabular*}{0.43\textwidth}{ll @{\extracolsep{\fill}} cccc}
		\toprule
		&$\ket{F'_1,F'}$ & $\ket{2,5/2}$ & $\ket{2,3/2}$ & $\ket{1,3/2}$ & $\ket{1,1/2}$\\
		$\ket{G,F_1,F}$&&&&&\\
		\midrule
		$\ket{1,1,3/2}$ && 13.52 &  0.44 &  1.73 &  3.68 \\
		$\ket{1,2,5/2}$ && 36.86 &  3.62 &  6.39 &  0.00 \\
		$\ket{1,1,1/2}$ &&  0.00 & 10.42 &  1.04 &  1.89 \\
		$\ket{1,2,3/2}$ &&  0.64 & 35.35 &  2.08 &  4.39 \\
		$\ket{1,0,1/2}$ &&  0.00 &  0.10 &  4.79 &  5.83 \\
		\addlinespace
		$\ket{2,1,1/2}$ &&  0.00 & 18.09 &  3.64 & 13.90 \\
		$\ket{2,2,3/2}$ &&  2.18 &  6.32 &  1.12 & 70.22 \\
		$\ket{2,1,3/2}$ && 21.13 &  1.20 & 22.72 &  0.08 \\
		$\ket{2,2,5/2}$ &&  4.14 &  0.22 & 53.55 &  0.00 \\
		$\ket{2,3,5/2}$ &&  0.09 & 24.23 &  2.95 &  0.00 \\
		$\ket{2,3,7/2}$ && 21.43 &  0.00 &  0.00 &  0.00 \\
		\bottomrule
	\end{tabular*}
	\caption{Branching ratios $|\braket{F'_1,F' || \mathrm{T}^1_p(\Op{d}) || G,F_1,F}|^2$
		corresponding to the squared reduced matrix elements of the electric dipole operator for the $A^2\Pi_{1/2}$, $J'=1/2^+$ state to the \gs, $N = 1$ state. Note, that these states are mixed eigenstates. In particular, $F_1$ is not a good quantum number and causes a violation of the selection rules $\Delta F_1 = 0,\pm 1$.
	}
	\label{tab:branchingratios}
\end{table}

\bibliography{biblio}

\begin{thebibliography}{69}%
\makeatletter
\providecommand \@ifxundefined [1]{%
 \@ifx{#1\undefined}
}%
\providecommand \@ifnum [1]{%
 \ifnum #1\expandafter \@firstoftwo
 \else \expandafter \@secondoftwo
 \fi
}%
\providecommand \@ifx [1]{%
 \ifx #1\expandafter \@firstoftwo
 \else \expandafter \@secondoftwo
 \fi
}%
\providecommand \natexlab [1]{#1}%
\providecommand \enquote  [1]{``#1''}%
\providecommand \bibnamefont  [1]{#1}%
\providecommand \bibfnamefont [1]{#1}%
\providecommand \citenamefont [1]{#1}%
\providecommand \href@noop [0]{\@secondoftwo}%
\providecommand \href [0]{\begingroup \@sanitize@url \@href}%
\providecommand \@href[1]{\@@startlink{#1}\@@href}%
\providecommand \@@href[1]{\endgroup#1\@@endlink}%
\providecommand \@sanitize@url [0]{\catcode `\\12\catcode `\$12\catcode
  `\&12\catcode `\#12\catcode `\^12\catcode `\_12\catcode `\%12\relax}%
\providecommand \@@startlink[1]{}%
\providecommand \@@endlink[0]{}%
\providecommand \url  [0]{\begingroup\@sanitize@url \@url }%
\providecommand \@url [1]{\endgroup\@href {#1}{\urlprefix }}%
\providecommand \urlprefix  [0]{URL }%
\providecommand \Eprint [0]{\href }%
\providecommand \doibase [0]{https://doi.org/}%
\providecommand \selectlanguage [0]{\@gobble}%
\providecommand \bibinfo  [0]{\@secondoftwo}%
\providecommand \bibfield  [0]{\@secondoftwo}%
\providecommand \translation [1]{[#1]}%
\providecommand \BibitemOpen [0]{}%
\providecommand \bibitemStop [0]{}%
\providecommand \bibitemNoStop [0]{.\EOS\space}%
\providecommand \EOS [0]{\spacefactor3000\relax}%
\providecommand \BibitemShut  [1]{\csname bibitem#1\endcsname}%
\let\auto@bib@innerbib\@empty
\bibitem [{\citenamefont {Carr}\ \emph {et~al.}(2009)\citenamefont {Carr},
  \citenamefont {DeMille}, \citenamefont {Krems},\ and\ \citenamefont
  {Ye}}]{Carr2009}%
  \BibitemOpen
  \bibfield  {author} {\bibinfo {author} {\bibfnamefont {L.~D.}\ \bibnamefont
  {Carr}}, \bibinfo {author} {\bibfnamefont {D.}~\bibnamefont {DeMille}},
  \bibinfo {author} {\bibfnamefont {R.~V.}\ \bibnamefont {Krems}},\ and\
  \bibinfo {author} {\bibfnamefont {J.}~\bibnamefont {Ye}},\ }\bibfield
  {title} {\bibinfo {title} {Cold and ultracold molecules: science, technology
  and applications},\ }\href {https://doi.org/10.1088/1367-2630/11/5/055049}
  {\bibfield  {journal} {\bibinfo  {journal} {New Journal of Physics}\ }\textbf
  {\bibinfo {volume} {11}},\ \bibinfo {pages} {055049} (\bibinfo {year}
  {2009})}\BibitemShut {NoStop}%
\bibitem [{\citenamefont {Bohn}\ \emph {et~al.}(2017)\citenamefont {Bohn},
  \citenamefont {Rey},\ and\ \citenamefont {Ye}}]{Bohn2017}%
  \BibitemOpen
  \bibfield  {author} {\bibinfo {author} {\bibfnamefont {J.~L.}\ \bibnamefont
  {Bohn}}, \bibinfo {author} {\bibfnamefont {A.~M.}\ \bibnamefont {Rey}},\ and\
  \bibinfo {author} {\bibfnamefont {J.}~\bibnamefont {Ye}},\ }\bibfield
  {title} {\bibinfo {title} {Cold molecules: Progress in quantum engineering of
  chemistry and quantum matter},\ }\href
  {https://doi.org/10.1126/science.aam6299} {\bibfield  {journal} {\bibinfo
  {journal} {Science}\ }\textbf {\bibinfo {volume} {357}},\ \bibinfo {pages}
  {1002} (\bibinfo {year} {2017})}\BibitemShut {NoStop}%
\bibitem [{\citenamefont {DeMille}\ \emph {et~al.}(2017)\citenamefont
  {DeMille}, \citenamefont {Doyle},\ and\ \citenamefont
  {Sushkov}}]{DeMille2017}%
  \BibitemOpen
  \bibfield  {author} {\bibinfo {author} {\bibfnamefont {D.}~\bibnamefont
  {DeMille}}, \bibinfo {author} {\bibfnamefont {J.~M.}\ \bibnamefont {Doyle}},\
  and\ \bibinfo {author} {\bibfnamefont {A.~O.}\ \bibnamefont {Sushkov}},\
  }\bibfield  {title} {\bibinfo {title} {{Probing the frontiers of particle
  physics with tabletop-scale experiments}},\ }\href
  {https://doi.org/10.1126/science.aal3003} {\bibfield  {journal} {\bibinfo
  {journal} {Science}\ }\textbf {\bibinfo {volume} {357}},\ \bibinfo {pages}
  {990} (\bibinfo {year} {2017})}\BibitemShut {NoStop}%
\bibitem [{\citenamefont {Safronova}\ \emph {et~al.}(2018)\citenamefont
  {Safronova}, \citenamefont {Budker}, \citenamefont {DeMille}, \citenamefont
  {Kimball}, \citenamefont {Derevianko},\ and\ \citenamefont
  {Clark}}]{Safronova2018}%
  \BibitemOpen
  \bibfield  {author} {\bibinfo {author} {\bibfnamefont {M.~S.}\ \bibnamefont
  {Safronova}}, \bibinfo {author} {\bibfnamefont {D.}~\bibnamefont {Budker}},
  \bibinfo {author} {\bibfnamefont {D.}~\bibnamefont {DeMille}}, \bibinfo
  {author} {\bibfnamefont {D.~F.~J.}\ \bibnamefont {Kimball}}, \bibinfo
  {author} {\bibfnamefont {A.}~\bibnamefont {Derevianko}},\ and\ \bibinfo
  {author} {\bibfnamefont {C.~W.}\ \bibnamefont {Clark}},\ }\bibfield  {title}
  {\bibinfo {title} {Search for new physics with atoms and molecules},\ }\href
  {https://doi.org/10.1103/RevModPhys.90.025008} {\bibfield  {journal}
  {\bibinfo  {journal} {Rev. Mod. Phys.}\ }\textbf {\bibinfo {volume} {90}},\
  \bibinfo {pages} {025008} (\bibinfo {year} {2018})}\BibitemShut {NoStop}%
\bibitem [{\citenamefont {Fitch}\ and\ \citenamefont
  {Tarbutt}(2021)}]{Fitch2021}%
  \BibitemOpen
  \bibfield  {author} {\bibinfo {author} {\bibfnamefont {N.}~\bibnamefont
  {Fitch}}\ and\ \bibinfo {author} {\bibfnamefont {M.}~\bibnamefont
  {Tarbutt}},\ }\bibfield  {title} {\bibinfo {title} {Laser cooled molecules},\
  }\href@noop {} {\bibfield  {journal} {\bibinfo  {journal} {arXiv:2103.00968}\
  } (\bibinfo {year} {2021})}\BibitemShut {NoStop}%
\bibitem [{\citenamefont {Barry}\ \emph {et~al.}(2014)\citenamefont {Barry},
  \citenamefont {McCarron}, \citenamefont {Norrgard}, \citenamefont
  {Steinecker},\ and\ \citenamefont {DeMille}}]{Barry2014}%
  \BibitemOpen
  \bibfield  {author} {\bibinfo {author} {\bibfnamefont {J.~F.}\ \bibnamefont
  {Barry}}, \bibinfo {author} {\bibfnamefont {D.~J.}\ \bibnamefont {McCarron}},
  \bibinfo {author} {\bibfnamefont {E.~B.}\ \bibnamefont {Norrgard}}, \bibinfo
  {author} {\bibfnamefont {M.~H.}\ \bibnamefont {Steinecker}},\ and\ \bibinfo
  {author} {\bibfnamefont {D.}~\bibnamefont {DeMille}},\ }\bibfield  {title}
  {\bibinfo {title} {{Magneto-optical trapping of a diatomic molecule}},\
  }\href {https://doi.org/10.1038/nature13634} {\bibfield  {journal} {\bibinfo
  {journal} {Nature}\ }\textbf {\bibinfo {volume} {512}},\ \bibinfo {pages}
  {286} (\bibinfo {year} {2014})}\BibitemShut {NoStop}%
\bibitem [{\citenamefont {Truppe}\ \emph {et~al.}(2017)\citenamefont {Truppe},
  \citenamefont {Williams}, \citenamefont {Hambach}, \citenamefont {Caldwell},
  \citenamefont {Fitch}, \citenamefont {Hinds}, \citenamefont {Sauer},\ and\
  \citenamefont {Tarbutt}}]{Truppe2017}%
  \BibitemOpen
  \bibfield  {author} {\bibinfo {author} {\bibfnamefont {S.}~\bibnamefont
  {Truppe}}, \bibinfo {author} {\bibfnamefont {H.~J.}\ \bibnamefont
  {Williams}}, \bibinfo {author} {\bibfnamefont {M.}~\bibnamefont {Hambach}},
  \bibinfo {author} {\bibfnamefont {L.}~\bibnamefont {Caldwell}}, \bibinfo
  {author} {\bibfnamefont {N.~J.}\ \bibnamefont {Fitch}}, \bibinfo {author}
  {\bibfnamefont {E.~A.}\ \bibnamefont {Hinds}}, \bibinfo {author}
  {\bibfnamefont {B.~E.}\ \bibnamefont {Sauer}},\ and\ \bibinfo {author}
  {\bibfnamefont {M.~R.}\ \bibnamefont {Tarbutt}},\ }\bibfield  {title}
  {\bibinfo {title} {{Molecules cooled below the Doppler limit}},\ }\href
  {https://doi.org/10.1038/nphys4241} {\bibfield  {journal} {\bibinfo
  {journal} {Nature Physics}\ }\textbf {\bibinfo {volume} {13}},\ \bibinfo
  {pages} {1173} (\bibinfo {year} {2017})}\BibitemShut {NoStop}%
\bibitem [{\citenamefont {Anderegg}\ \emph {et~al.}(2017)\citenamefont
  {Anderegg}, \citenamefont {Augenbraun}, \citenamefont {Chae}, \citenamefont
  {Hemmerling}, \citenamefont {Hutzler}, \citenamefont {Ravi}, \citenamefont
  {Collopy}, \citenamefont {Ye}, \citenamefont {Ketterle},\ and\ \citenamefont
  {Doyle}}]{Anderegg2017}%
  \BibitemOpen
  \bibfield  {author} {\bibinfo {author} {\bibfnamefont {L.}~\bibnamefont
  {Anderegg}}, \bibinfo {author} {\bibfnamefont {B.~L.}\ \bibnamefont
  {Augenbraun}}, \bibinfo {author} {\bibfnamefont {E.}~\bibnamefont {Chae}},
  \bibinfo {author} {\bibfnamefont {B.}~\bibnamefont {Hemmerling}}, \bibinfo
  {author} {\bibfnamefont {N.~R.}\ \bibnamefont {Hutzler}}, \bibinfo {author}
  {\bibfnamefont {A.}~\bibnamefont {Ravi}}, \bibinfo {author} {\bibfnamefont
  {A.}~\bibnamefont {Collopy}}, \bibinfo {author} {\bibfnamefont
  {J.}~\bibnamefont {Ye}}, \bibinfo {author} {\bibfnamefont {W.}~\bibnamefont
  {Ketterle}},\ and\ \bibinfo {author} {\bibfnamefont {J.~M.}\ \bibnamefont
  {Doyle}},\ }\bibfield  {title} {\bibinfo {title} {{Radio Frequency
  Magneto-Optical Trapping of CaF with High Density}},\ }\href
  {https://doi.org/10.1103/PhysRevLett.119.103201} {\bibfield  {journal}
  {\bibinfo  {journal} {Phys. Rev. Lett.}\ }\textbf {\bibinfo {volume} {119}},\
  \bibinfo {pages} {103201} (\bibinfo {year} {2017})}\BibitemShut {NoStop}%
\bibitem [{\citenamefont {Collopy}\ \emph {et~al.}(2018)\citenamefont
  {Collopy}, \citenamefont {Ding}, \citenamefont {Wu}, \citenamefont
  {Finneran}, \citenamefont {Anderegg}, \citenamefont {Augenbraun},
  \citenamefont {Doyle},\ and\ \citenamefont {Ye}}]{Collopy2018}%
  \BibitemOpen
  \bibfield  {author} {\bibinfo {author} {\bibfnamefont {A.~L.}\ \bibnamefont
  {Collopy}}, \bibinfo {author} {\bibfnamefont {S.}~\bibnamefont {Ding}},
  \bibinfo {author} {\bibfnamefont {Y.}~\bibnamefont {Wu}}, \bibinfo {author}
  {\bibfnamefont {I.~A.}\ \bibnamefont {Finneran}}, \bibinfo {author}
  {\bibfnamefont {L.}~\bibnamefont {Anderegg}}, \bibinfo {author}
  {\bibfnamefont {B.~L.}\ \bibnamefont {Augenbraun}}, \bibinfo {author}
  {\bibfnamefont {J.~M.}\ \bibnamefont {Doyle}},\ and\ \bibinfo {author}
  {\bibfnamefont {J.}~\bibnamefont {Ye}},\ }\bibfield  {title} {\bibinfo
  {title} {{3D Magneto-Optical Trap of Yttrium Monoxide}},\ }\href
  {https://doi.org/10.1103/PhysRevLett.121.213201} {\bibfield  {journal}
  {\bibinfo  {journal} {Phys. Rev. Lett.}\ }\textbf {\bibinfo {volume} {121}},\
  \bibinfo {pages} {213201} (\bibinfo {year} {2018})}\BibitemShut {NoStop}%
\bibitem [{\citenamefont {Anderegg}\ \emph {et~al.}(2018)\citenamefont
  {Anderegg}, \citenamefont {Augenbraun}, \citenamefont {Bao}, \citenamefont
  {Burchesky}, \citenamefont {Cheuk}, \citenamefont {Ketterle},\ and\
  \citenamefont {Doyle}}]{Anderegg2018}%
  \BibitemOpen
  \bibfield  {author} {\bibinfo {author} {\bibfnamefont {L.}~\bibnamefont
  {Anderegg}}, \bibinfo {author} {\bibfnamefont {B.~L.}\ \bibnamefont
  {Augenbraun}}, \bibinfo {author} {\bibfnamefont {Y.}~\bibnamefont {Bao}},
  \bibinfo {author} {\bibfnamefont {S.}~\bibnamefont {Burchesky}}, \bibinfo
  {author} {\bibfnamefont {L.~W.}\ \bibnamefont {Cheuk}}, \bibinfo {author}
  {\bibfnamefont {W.}~\bibnamefont {Ketterle}},\ and\ \bibinfo {author}
  {\bibfnamefont {J.~M.}\ \bibnamefont {Doyle}},\ }\bibfield  {title} {\bibinfo
  {title} {{Laser cooling of optically trapped molecules}},\ }\href
  {https://doi.org/10.1038/s41567-018-0191-z} {\bibfield  {journal} {\bibinfo
  {journal} {Nature Physics}\ }\textbf {\bibinfo {volume} {14}},\ \bibinfo
  {pages} {890} (\bibinfo {year} {2018})}\BibitemShut {NoStop}%
\bibitem [{\citenamefont {McCarron}\ \emph {et~al.}(2018)\citenamefont
  {McCarron}, \citenamefont {Steinecker}, \citenamefont {Zhu},\ and\
  \citenamefont {DeMille}}]{McCarron2018}%
  \BibitemOpen
  \bibfield  {author} {\bibinfo {author} {\bibfnamefont {D.~J.}\ \bibnamefont
  {McCarron}}, \bibinfo {author} {\bibfnamefont {M.~H.}\ \bibnamefont
  {Steinecker}}, \bibinfo {author} {\bibfnamefont {Y.}~\bibnamefont {Zhu}},\
  and\ \bibinfo {author} {\bibfnamefont {D.}~\bibnamefont {DeMille}},\
  }\bibfield  {title} {\bibinfo {title} {Magnetic trapping of an ultracold gas
  of polar molecules},\ }\href {https://doi.org/10.1103/PhysRevLett.121.013202}
  {\bibfield  {journal} {\bibinfo  {journal} {Phys. Rev. Lett.}\ }\textbf
  {\bibinfo {volume} {121}},\ \bibinfo {pages} {013202} (\bibinfo {year}
  {2018})}\BibitemShut {NoStop}%
\bibitem [{\citenamefont {Ding}\ \emph {et~al.}(2020)\citenamefont {Ding},
  \citenamefont {Wu}, \citenamefont {Finneran}, \citenamefont {Burau},\ and\
  \citenamefont {Ye}}]{Ding2020}%
  \BibitemOpen
  \bibfield  {author} {\bibinfo {author} {\bibfnamefont {S.}~\bibnamefont
  {Ding}}, \bibinfo {author} {\bibfnamefont {Y.}~\bibnamefont {Wu}}, \bibinfo
  {author} {\bibfnamefont {I.~A.}\ \bibnamefont {Finneran}}, \bibinfo {author}
  {\bibfnamefont {J.~J.}\ \bibnamefont {Burau}},\ and\ \bibinfo {author}
  {\bibfnamefont {J.}~\bibnamefont {Ye}},\ }\bibfield  {title} {\bibinfo
  {title} {Sub-doppler cooling and compressed trapping of yo molecules at
  $\ensuremath{\mu}\mathrm{K}$ temperatures},\ }\href
  {https://doi.org/10.1103/PhysRevX.10.021049} {\bibfield  {journal} {\bibinfo
  {journal} {Phys. Rev. X}\ }\textbf {\bibinfo {volume} {10}},\ \bibinfo
  {pages} {021049} (\bibinfo {year} {2020})}\BibitemShut {NoStop}%
\bibitem [{\citenamefont {Williams}\ \emph {et~al.}(2018)\citenamefont
  {Williams}, \citenamefont {Caldwell}, \citenamefont {Fitch}, \citenamefont
  {Truppe}, \citenamefont {Rodewald}, \citenamefont {Hinds}, \citenamefont
  {Sauer},\ and\ \citenamefont {Tarbutt}}]{Williams2018}%
  \BibitemOpen
  \bibfield  {author} {\bibinfo {author} {\bibfnamefont {H.~J.}\ \bibnamefont
  {Williams}}, \bibinfo {author} {\bibfnamefont {L.}~\bibnamefont {Caldwell}},
  \bibinfo {author} {\bibfnamefont {N.~J.}\ \bibnamefont {Fitch}}, \bibinfo
  {author} {\bibfnamefont {S.}~\bibnamefont {Truppe}}, \bibinfo {author}
  {\bibfnamefont {J.}~\bibnamefont {Rodewald}}, \bibinfo {author}
  {\bibfnamefont {E.~A.}\ \bibnamefont {Hinds}}, \bibinfo {author}
  {\bibfnamefont {B.~E.}\ \bibnamefont {Sauer}},\ and\ \bibinfo {author}
  {\bibfnamefont {M.~R.}\ \bibnamefont {Tarbutt}},\ }\bibfield  {title}
  {\bibinfo {title} {Magnetic trapping and coherent control of laser-cooled
  molecules},\ }\href {https://doi.org/10.1103/PhysRevLett.120.163201}
  {\bibfield  {journal} {\bibinfo  {journal} {Phys. Rev. Lett.}\ }\textbf
  {\bibinfo {volume} {120}},\ \bibinfo {pages} {163201} (\bibinfo {year}
  {2018})}\BibitemShut {NoStop}%
\bibitem [{\citenamefont {Anderegg}\ \emph {et~al.}(2019)\citenamefont
  {Anderegg}, \citenamefont {Cheuk}, \citenamefont {Bao}, \citenamefont
  {Burchesky}, \citenamefont {Ketterle}, \citenamefont {Ni},\ and\
  \citenamefont {Doyle}}]{Anderegg2019}%
  \BibitemOpen
  \bibfield  {author} {\bibinfo {author} {\bibfnamefont {L.}~\bibnamefont
  {Anderegg}}, \bibinfo {author} {\bibfnamefont {L.~W.}\ \bibnamefont {Cheuk}},
  \bibinfo {author} {\bibfnamefont {Y.}~\bibnamefont {Bao}}, \bibinfo {author}
  {\bibfnamefont {S.}~\bibnamefont {Burchesky}}, \bibinfo {author}
  {\bibfnamefont {W.}~\bibnamefont {Ketterle}}, \bibinfo {author}
  {\bibfnamefont {K.-K.}\ \bibnamefont {Ni}},\ and\ \bibinfo {author}
  {\bibfnamefont {J.~M.}\ \bibnamefont {Doyle}},\ }\bibfield  {title} {\bibinfo
  {title} {{An optical tweezer array of ultracold molecules}},\ }\href
  {https://doi.org/10.1126/science.aax1265} {\bibfield  {journal} {\bibinfo
  {journal} {Science}\ }\textbf {\bibinfo {volume} {365}},\ \bibinfo {pages}
  {1156} (\bibinfo {year} {2019})}\BibitemShut {NoStop}%
\bibitem [{\citenamefont {Cheuk}\ \emph {et~al.}(2020)\citenamefont {Cheuk},
  \citenamefont {Anderegg}, \citenamefont {Bao}, \citenamefont {Burchesky},
  \citenamefont {Yu}, \citenamefont {Ketterle}, \citenamefont {Ni},\ and\
  \citenamefont {Doyle}}]{Cheuk2020}%
  \BibitemOpen
  \bibfield  {author} {\bibinfo {author} {\bibfnamefont {L.~W.}\ \bibnamefont
  {Cheuk}}, \bibinfo {author} {\bibfnamefont {L.}~\bibnamefont {Anderegg}},
  \bibinfo {author} {\bibfnamefont {Y.}~\bibnamefont {Bao}}, \bibinfo {author}
  {\bibfnamefont {S.}~\bibnamefont {Burchesky}}, \bibinfo {author}
  {\bibfnamefont {S.~S.}\ \bibnamefont {Yu}}, \bibinfo {author} {\bibfnamefont
  {W.}~\bibnamefont {Ketterle}}, \bibinfo {author} {\bibfnamefont {K.-K.}\
  \bibnamefont {Ni}},\ and\ \bibinfo {author} {\bibfnamefont {J.~M.}\
  \bibnamefont {Doyle}},\ }\bibfield  {title} {\bibinfo {title} {{Observation
  of Collisions between Two Ultracold Ground-State CaF Molecules}},\ }\href
  {https://doi.org/10.1103/PhysRevLett.125.043401} {\bibfield  {journal}
  {\bibinfo  {journal} {Phys. Rev. Lett.}\ }\textbf {\bibinfo {volume} {125}},\
  \bibinfo {pages} {43401} (\bibinfo {year} {2020})}\BibitemShut {NoStop}%
\bibitem [{\citenamefont {Chen}\ \emph {et~al.}(2017)\citenamefont {Chen},
  \citenamefont {Bu},\ and\ \citenamefont {Yan}}]{Chen2017}%
  \BibitemOpen
  \bibfield  {author} {\bibinfo {author} {\bibfnamefont {T.}~\bibnamefont
  {Chen}}, \bibinfo {author} {\bibfnamefont {W.}~\bibnamefont {Bu}},\ and\
  \bibinfo {author} {\bibfnamefont {B.}~\bibnamefont {Yan}},\ }\bibfield
  {title} {\bibinfo {title} {Radiative deflection of a baf molecular beam via
  optical cycling},\ }\href {https://doi.org/10.1103/PhysRevA.96.053401}
  {\bibfield  {journal} {\bibinfo  {journal} {Phys. Rev. A}\ }\textbf {\bibinfo
  {volume} {96}},\ \bibinfo {pages} {053401} (\bibinfo {year}
  {2017})}\BibitemShut {NoStop}%
\bibitem [{\citenamefont {Norrgard}\ \emph {et~al.}(2017)\citenamefont
  {Norrgard}, \citenamefont {Edwards}, \citenamefont {McCarron}, \citenamefont
  {Steinecker}, \citenamefont {DeMille}, \citenamefont {Alam}, \citenamefont
  {Peck}, \citenamefont {Wadia},\ and\ \citenamefont {Hunter}}]{Norrgard2017}%
  \BibitemOpen
  \bibfield  {author} {\bibinfo {author} {\bibfnamefont {E.~B.}\ \bibnamefont
  {Norrgard}}, \bibinfo {author} {\bibfnamefont {E.~R.}\ \bibnamefont
  {Edwards}}, \bibinfo {author} {\bibfnamefont {D.~J.}\ \bibnamefont
  {McCarron}}, \bibinfo {author} {\bibfnamefont {M.~H.}\ \bibnamefont
  {Steinecker}}, \bibinfo {author} {\bibfnamefont {D.}~\bibnamefont {DeMille}},
  \bibinfo {author} {\bibfnamefont {S.~S.}\ \bibnamefont {Alam}}, \bibinfo
  {author} {\bibfnamefont {S.~K.}\ \bibnamefont {Peck}}, \bibinfo {author}
  {\bibfnamefont {N.~S.}\ \bibnamefont {Wadia}},\ and\ \bibinfo {author}
  {\bibfnamefont {L.~R.}\ \bibnamefont {Hunter}},\ }\bibfield  {title}
  {\bibinfo {title} {Hyperfine structure of the
  $b{\phantom{\rule{0.16em}{0ex}}}^{3}{\mathrm{\ensuremath{\Pi}}}_{1}$ state
  and predictions of optical cycling behavior in the
  $x\ensuremath{\rightarrow}b$ transition of {TlF}},\ }\href
  {https://doi.org/10.1103/PhysRevA.95.062506} {\bibfield  {journal} {\bibinfo
  {journal} {Phys. Rev. A}\ }\textbf {\bibinfo {volume} {95}},\ \bibinfo
  {pages} {062506} (\bibinfo {year} {2017})}\BibitemShut {NoStop}%
\bibitem [{\citenamefont {Lim}\ \emph {et~al.}(2018)\citenamefont {Lim},
  \citenamefont {Almond}, \citenamefont {Trigatzis}, \citenamefont {Devlin},
  \citenamefont {Fitch}, \citenamefont {Sauer}, \citenamefont {Tarbutt},\ and\
  \citenamefont {Hinds}}]{Lim2018}%
  \BibitemOpen
  \bibfield  {author} {\bibinfo {author} {\bibfnamefont {J.}~\bibnamefont
  {Lim}}, \bibinfo {author} {\bibfnamefont {J.~R.}\ \bibnamefont {Almond}},
  \bibinfo {author} {\bibfnamefont {M.~A.}\ \bibnamefont {Trigatzis}}, \bibinfo
  {author} {\bibfnamefont {J.~A.}\ \bibnamefont {Devlin}}, \bibinfo {author}
  {\bibfnamefont {N.~J.}\ \bibnamefont {Fitch}}, \bibinfo {author}
  {\bibfnamefont {B.~E.}\ \bibnamefont {Sauer}}, \bibinfo {author}
  {\bibfnamefont {M.~R.}\ \bibnamefont {Tarbutt}},\ and\ \bibinfo {author}
  {\bibfnamefont {E.~A.}\ \bibnamefont {Hinds}},\ }\bibfield  {title} {\bibinfo
  {title} {Laser cooled ybf molecules for measuring the electron's electric
  dipole moment},\ }\href {https://doi.org/10.1103/PhysRevLett.120.123201}
  {\bibfield  {journal} {\bibinfo  {journal} {Phys. Rev. Lett.}\ }\textbf
  {\bibinfo {volume} {120}},\ \bibinfo {pages} {123201} (\bibinfo {year}
  {2018})}\BibitemShut {NoStop}%
\bibitem [{\citenamefont {Truppe}\ \emph {et~al.}(2019)\citenamefont {Truppe},
  \citenamefont {Marx}, \citenamefont {Kray}, \citenamefont {Doppelbauer},
  \citenamefont {Hofs\"ass}, \citenamefont {Schewe}, \citenamefont {Walter},
  \citenamefont {P\'erez-R\'{\i}os}, \citenamefont {Sartakov},\ and\
  \citenamefont {Meijer}}]{Truppe2019}%
  \BibitemOpen
  \bibfield  {author} {\bibinfo {author} {\bibfnamefont {S.}~\bibnamefont
  {Truppe}}, \bibinfo {author} {\bibfnamefont {S.}~\bibnamefont {Marx}},
  \bibinfo {author} {\bibfnamefont {S.}~\bibnamefont {Kray}}, \bibinfo {author}
  {\bibfnamefont {M.}~\bibnamefont {Doppelbauer}}, \bibinfo {author}
  {\bibfnamefont {S.}~\bibnamefont {Hofs\"ass}}, \bibinfo {author}
  {\bibfnamefont {H.~C.}\ \bibnamefont {Schewe}}, \bibinfo {author}
  {\bibfnamefont {N.}~\bibnamefont {Walter}}, \bibinfo {author} {\bibfnamefont
  {J.}~\bibnamefont {P\'erez-R\'{\i}os}}, \bibinfo {author} {\bibfnamefont
  {B.~G.}\ \bibnamefont {Sartakov}},\ and\ \bibinfo {author} {\bibfnamefont
  {G.}~\bibnamefont {Meijer}},\ }\bibfield  {title} {\bibinfo {title}
  {Spectroscopic characterization of aluminum monofluoride with relevance to
  laser cooling and trapping},\ }\href
  {https://doi.org/10.1103/PhysRevA.100.052513} {\bibfield  {journal} {\bibinfo
   {journal} {Phys. Rev. A}\ }\textbf {\bibinfo {volume} {100}},\ \bibinfo
  {pages} {052513} (\bibinfo {year} {2019})}\BibitemShut {NoStop}%
\bibitem [{\citenamefont {McNally}\ \emph {et~al.}(2020)\citenamefont
  {McNally}, \citenamefont {Kozyryev}, \citenamefont {Vazquez-Carson},
  \citenamefont {Wenz}, \citenamefont {Wang},\ and\ \citenamefont
  {Zelevinsky}}]{McNally2020}%
  \BibitemOpen
  \bibfield  {author} {\bibinfo {author} {\bibfnamefont {R.~L.}\ \bibnamefont
  {McNally}}, \bibinfo {author} {\bibfnamefont {I.}~\bibnamefont {Kozyryev}},
  \bibinfo {author} {\bibfnamefont {S.}~\bibnamefont {Vazquez-Carson}},
  \bibinfo {author} {\bibfnamefont {K.}~\bibnamefont {Wenz}}, \bibinfo {author}
  {\bibfnamefont {T.}~\bibnamefont {Wang}},\ and\ \bibinfo {author}
  {\bibfnamefont {T.}~\bibnamefont {Zelevinsky}},\ }\bibfield  {title}
  {\bibinfo {title} {Optical cycling, radiative deflection and laser cooling of
  barium monohydride (${}^{138}${BaH})},\ }\href
  {https://doi.org/10.1088/1367-2630/aba3e9} {\bibfield  {journal} {\bibinfo
  {journal} {New Journal of Physics}\ }\textbf {\bibinfo {volume} {22}},\
  \bibinfo {pages} {083047} (\bibinfo {year} {2020})}\BibitemShut {NoStop}%
\bibitem [{\citenamefont {Albrecht}\ \emph {et~al.}(2020)\citenamefont
  {Albrecht}, \citenamefont {Scharwaechter}, \citenamefont {Sixt},
  \citenamefont {Hofer},\ and\ \citenamefont {Langen}}]{Albrecht2020}%
  \BibitemOpen
  \bibfield  {author} {\bibinfo {author} {\bibfnamefont {R.}~\bibnamefont
  {Albrecht}}, \bibinfo {author} {\bibfnamefont {M.}~\bibnamefont
  {Scharwaechter}}, \bibinfo {author} {\bibfnamefont {T.}~\bibnamefont {Sixt}},
  \bibinfo {author} {\bibfnamefont {L.}~\bibnamefont {Hofer}},\ and\ \bibinfo
  {author} {\bibfnamefont {T.}~\bibnamefont {Langen}},\ }\bibfield  {title}
  {\bibinfo {title} {Buffer-gas cooling, high-resolution spectroscopy, and
  optical cycling of barium monofluoride molecules},\ }\href
  {https://doi.org/10.1103/PhysRevA.101.013413} {\bibfield  {journal} {\bibinfo
   {journal} {Phys. Rev. A}\ }\textbf {\bibinfo {volume} {101}},\ \bibinfo
  {pages} {013413} (\bibinfo {year} {2020})}\BibitemShut {NoStop}%
\bibitem [{\citenamefont {Isaev}\ and\ \citenamefont
  {Berger}(2016)}]{Isaev2016}%
  \BibitemOpen
  \bibfield  {author} {\bibinfo {author} {\bibfnamefont {T.~A.}\ \bibnamefont
  {Isaev}}\ and\ \bibinfo {author} {\bibfnamefont {R.}~\bibnamefont {Berger}},\
  }\bibfield  {title} {\bibinfo {title} {Polyatomic candidates for cooling of
  molecules with lasers from simple theoretical concepts},\ }\href
  {https://doi.org/10.1103/PhysRevLett.116.063006} {\bibfield  {journal}
  {\bibinfo  {journal} {Phys. Rev. Lett.}\ }\textbf {\bibinfo {volume} {116}},\
  \bibinfo {pages} {063006} (\bibinfo {year} {2016})}\BibitemShut {NoStop}%
\bibitem [{\citenamefont {Kozyryev}\ \emph {et~al.}(2017)\citenamefont
  {Kozyryev}, \citenamefont {Baum}, \citenamefont {Matsuda}, \citenamefont
  {Augenbraun}, \citenamefont {Anderegg}, \citenamefont {Sedlack},\ and\
  \citenamefont {Doyle}}]{Kozyryev2017}%
  \BibitemOpen
  \bibfield  {author} {\bibinfo {author} {\bibfnamefont {I.}~\bibnamefont
  {Kozyryev}}, \bibinfo {author} {\bibfnamefont {L.}~\bibnamefont {Baum}},
  \bibinfo {author} {\bibfnamefont {K.}~\bibnamefont {Matsuda}}, \bibinfo
  {author} {\bibfnamefont {B.~L.}\ \bibnamefont {Augenbraun}}, \bibinfo
  {author} {\bibfnamefont {L.}~\bibnamefont {Anderegg}}, \bibinfo {author}
  {\bibfnamefont {A.~P.}\ \bibnamefont {Sedlack}},\ and\ \bibinfo {author}
  {\bibfnamefont {J.~M.}\ \bibnamefont {Doyle}},\ }\bibfield  {title} {\bibinfo
  {title} {Sisyphus laser cooling of a polyatomic molecule},\ }\href
  {https://doi.org/10.1103/PhysRevLett.118.173201} {\bibfield  {journal}
  {\bibinfo  {journal} {Phys. Rev. Lett.}\ }\textbf {\bibinfo {volume} {118}},\
  \bibinfo {pages} {173201} (\bibinfo {year} {2017})}\BibitemShut {NoStop}%
\bibitem [{\citenamefont {Baum}\ \emph {et~al.}(2020)\citenamefont {Baum},
  \citenamefont {Vilas}, \citenamefont {Hallas}, \citenamefont {Augenbraun},
  \citenamefont {Raval}, \citenamefont {Mitra},\ and\ \citenamefont
  {Doyle}}]{Baum2020}%
  \BibitemOpen
  \bibfield  {author} {\bibinfo {author} {\bibfnamefont {L.}~\bibnamefont
  {Baum}}, \bibinfo {author} {\bibfnamefont {N.~B.}\ \bibnamefont {Vilas}},
  \bibinfo {author} {\bibfnamefont {C.}~\bibnamefont {Hallas}}, \bibinfo
  {author} {\bibfnamefont {B.~L.}\ \bibnamefont {Augenbraun}}, \bibinfo
  {author} {\bibfnamefont {S.}~\bibnamefont {Raval}}, \bibinfo {author}
  {\bibfnamefont {D.}~\bibnamefont {Mitra}},\ and\ \bibinfo {author}
  {\bibfnamefont {J.~M.}\ \bibnamefont {Doyle}},\ }\bibfield  {title} {\bibinfo
  {title} {1d magneto-optical trap of polyatomic molecules},\ }\href
  {https://doi.org/10.1103/PhysRevLett.124.133201} {\bibfield  {journal}
  {\bibinfo  {journal} {Phys. Rev. Lett.}\ }\textbf {\bibinfo {volume} {124}},\
  \bibinfo {pages} {133201} (\bibinfo {year} {2020})}\BibitemShut {NoStop}%
\bibitem [{\citenamefont {Mitra}\ \emph {et~al.}(2020)\citenamefont {Mitra},
  \citenamefont {Vilas}, \citenamefont {Hallas}, \citenamefont {Anderegg},
  \citenamefont {Augenbraun}, \citenamefont {Baum}, \citenamefont {Miller},
  \citenamefont {Raval},\ and\ \citenamefont {Doyle}}]{Mitra2020}%
  \BibitemOpen
  \bibfield  {author} {\bibinfo {author} {\bibfnamefont {D.}~\bibnamefont
  {Mitra}}, \bibinfo {author} {\bibfnamefont {N.~B.}\ \bibnamefont {Vilas}},
  \bibinfo {author} {\bibfnamefont {C.}~\bibnamefont {Hallas}}, \bibinfo
  {author} {\bibfnamefont {L.}~\bibnamefont {Anderegg}}, \bibinfo {author}
  {\bibfnamefont {B.~L.}\ \bibnamefont {Augenbraun}}, \bibinfo {author}
  {\bibfnamefont {L.}~\bibnamefont {Baum}}, \bibinfo {author} {\bibfnamefont
  {C.}~\bibnamefont {Miller}}, \bibinfo {author} {\bibfnamefont
  {S.}~\bibnamefont {Raval}},\ and\ \bibinfo {author} {\bibfnamefont {J.~M.}\
  \bibnamefont {Doyle}},\ }\bibfield  {title} {\bibinfo {title} {Direct laser
  cooling of a symmetric top molecule},\ }\href
  {https://doi.org/10.1126/science.abc5357} {\bibfield  {journal} {\bibinfo
  {journal} {Science}\ }\textbf {\bibinfo {volume} {369}},\ \bibinfo {pages}
  {1366} (\bibinfo {year} {2020})}\BibitemShut {NoStop}%
\bibitem [{\citenamefont {Wood}\ \emph {et~al.}(1997)\citenamefont {Wood},
  \citenamefont {Bennett}, \citenamefont {Cho}, \citenamefont {Masterson},
  \citenamefont {Roberts}, \citenamefont {Tanner},\ and\ \citenamefont
  {Wieman}}]{Wood1997}%
  \BibitemOpen
  \bibfield  {author} {\bibinfo {author} {\bibfnamefont {C.~S.}\ \bibnamefont
  {Wood}}, \bibinfo {author} {\bibfnamefont {S.~C.}\ \bibnamefont {Bennett}},
  \bibinfo {author} {\bibfnamefont {D.}~\bibnamefont {Cho}}, \bibinfo {author}
  {\bibfnamefont {B.~P.}\ \bibnamefont {Masterson}}, \bibinfo {author}
  {\bibfnamefont {J.~L.}\ \bibnamefont {Roberts}}, \bibinfo {author}
  {\bibfnamefont {C.~E.}\ \bibnamefont {Tanner}},\ and\ \bibinfo {author}
  {\bibfnamefont {C.~E.}\ \bibnamefont {Wieman}},\ }\bibfield  {title}
  {\bibinfo {title} {Measurement of parity nonconservation and an anapole
  moment in cesium},\ }\href@noop {} {\bibfield  {journal} {\bibinfo  {journal}
  {Science}\ }\textbf {\bibinfo {volume} {275}},\ \bibinfo {pages} {1759}
  (\bibinfo {year} {1997})}\BibitemShut {NoStop}%
\bibitem [{\citenamefont {DeMille}\ \emph {et~al.}(2008)\citenamefont
  {DeMille}, \citenamefont {Cahn}, \citenamefont {Murphree}, \citenamefont
  {Rahmlow},\ and\ \citenamefont {Kozlov}}]{Demille2008}%
  \BibitemOpen
  \bibfield  {author} {\bibinfo {author} {\bibfnamefont {D.}~\bibnamefont
  {DeMille}}, \bibinfo {author} {\bibfnamefont {S.~B.}\ \bibnamefont {Cahn}},
  \bibinfo {author} {\bibfnamefont {D.}~\bibnamefont {Murphree}}, \bibinfo
  {author} {\bibfnamefont {D.~A.}\ \bibnamefont {Rahmlow}},\ and\ \bibinfo
  {author} {\bibfnamefont {M.~G.}\ \bibnamefont {Kozlov}},\ }\bibfield  {title}
  {\bibinfo {title} {{Using Molecules to Measure Nuclear Spin-Dependent Parity
  Violation}},\ }\href {https://doi.org/10.1103/PhysRevLett.100.023003}
  {\bibfield  {journal} {\bibinfo  {journal} {Phys. Rev. Lett.}\ }\textbf
  {\bibinfo {volume} {100}},\ \bibinfo {pages} {23003} (\bibinfo {year}
  {2008})}\BibitemShut {NoStop}%
\bibitem [{\citenamefont {Altuntas}\ \emph {et~al.}(2018)\citenamefont
  {Altuntas}, \citenamefont {Ammon}, \citenamefont {Cahn},\ and\ \citenamefont
  {DeMille}}]{Altuntas2018}%
  \BibitemOpen
  \bibfield  {author} {\bibinfo {author} {\bibfnamefont {E.}~\bibnamefont
  {Altuntas}}, \bibinfo {author} {\bibfnamefont {J.}~\bibnamefont {Ammon}},
  \bibinfo {author} {\bibfnamefont {S.~B.}\ \bibnamefont {Cahn}},\ and\
  \bibinfo {author} {\bibfnamefont {D.}~\bibnamefont {DeMille}},\ }\bibfield
  {title} {\bibinfo {title} {{Demonstration of a Sensitive Method to Measure
  Nuclear-Spin-Dependent Parity Violation}},\ }\href
  {https://doi.org/10.1103/PhysRevLett.120.142501} {\bibfield  {journal}
  {\bibinfo  {journal} {Phys. Rev. Lett.}\ }\textbf {\bibinfo {volume} {120}},\
  \bibinfo {pages} {142501} (\bibinfo {year} {2018})}\BibitemShut {NoStop}%
\bibitem [{\citenamefont {Aggarwal}\ \emph {et~al.}(2018)\citenamefont
  {Aggarwal}, \citenamefont {Bethlem}, \citenamefont {Borschevsky},
  \citenamefont {Denis}, \citenamefont {Esajas}, \citenamefont {Haase},
  \citenamefont {Hao}, \citenamefont {Hoekstra}, \citenamefont {Jungmann},
  \citenamefont {Meijknecht}, \citenamefont {Mooij}, \citenamefont
  {Timmermans}, \citenamefont {Ubachs}, \citenamefont {Willmann},\ and\
  \citenamefont {Zapara}}]{Aggarwal2018}%
  \BibitemOpen
  \bibfield  {author} {\bibinfo {author} {\bibfnamefont {P.}~\bibnamefont
  {Aggarwal}}, \bibinfo {author} {\bibfnamefont {H.~L.}\ \bibnamefont
  {Bethlem}}, \bibinfo {author} {\bibfnamefont {A.}~\bibnamefont
  {Borschevsky}}, \bibinfo {author} {\bibfnamefont {M.}~\bibnamefont {Denis}},
  \bibinfo {author} {\bibfnamefont {K.}~\bibnamefont {Esajas}}, \bibinfo
  {author} {\bibfnamefont {P.~A.~B.}\ \bibnamefont {Haase}}, \bibinfo {author}
  {\bibfnamefont {Y.}~\bibnamefont {Hao}}, \bibinfo {author} {\bibfnamefont
  {S.}~\bibnamefont {Hoekstra}}, \bibinfo {author} {\bibfnamefont
  {K.}~\bibnamefont {Jungmann}}, \bibinfo {author} {\bibfnamefont {T.~B.}\
  \bibnamefont {Meijknecht}}, \bibinfo {author} {\bibfnamefont {M.~C.}\
  \bibnamefont {Mooij}}, \bibinfo {author} {\bibfnamefont {R.~G.~E.}\
  \bibnamefont {Timmermans}}, \bibinfo {author} {\bibfnamefont
  {W.}~\bibnamefont {Ubachs}}, \bibinfo {author} {\bibfnamefont
  {L.}~\bibnamefont {Willmann}},\ and\ \bibinfo {author} {\bibfnamefont
  {A.}~\bibnamefont {Zapara}},\ }\bibfield  {title} {\bibinfo {title}
  {{Measuring the electric dipole moment of the electron in BaF}},\ }\href
  {https://doi.org/10.1140/epjd/e2018-90192-9} {\bibfield  {journal} {\bibinfo
  {journal} {Eur. Phys. J. D}\ }\textbf {\bibinfo {volume} {72}},\ \bibinfo
  {pages} {197} (\bibinfo {year} {2018})}\BibitemShut {NoStop}%
\bibitem [{\citenamefont {{Garcia Ruiz}}\ \emph {et~al.}(2020)\citenamefont
  {{Garcia Ruiz}}, \citenamefont {Berger}, \citenamefont {Billowes},
  \citenamefont {Binnersley}, \citenamefont {Bissell}, \citenamefont {Breier},
  \citenamefont {Brinson}, \citenamefont {Chrysalidis}, \citenamefont
  {Cocolios}, \citenamefont {Cooper}, \citenamefont {Flanagan}, \citenamefont
  {Giesen}, \citenamefont {de~Groote}, \citenamefont {Franchoo}, \citenamefont
  {Gustafsson}, \citenamefont {Isaev}, \citenamefont {Koszor{\'{u}}s},
  \citenamefont {Neyens}, \citenamefont {Perrett}, \citenamefont {Ricketts},
  \citenamefont {Rothe}, \citenamefont {Schweikhard}, \citenamefont {Vernon},
  \citenamefont {Wendt}, \citenamefont {Wienholtz}, \citenamefont {Wilkins},\
  and\ \citenamefont {Yang}}]{GarciaRuiz2020}%
  \BibitemOpen
  \bibfield  {author} {\bibinfo {author} {\bibfnamefont {R.~F.}\ \bibnamefont
  {{Garcia Ruiz}}}, \bibinfo {author} {\bibfnamefont {R.}~\bibnamefont
  {Berger}}, \bibinfo {author} {\bibfnamefont {J.}~\bibnamefont {Billowes}},
  \bibinfo {author} {\bibfnamefont {C.~L.}\ \bibnamefont {Binnersley}},
  \bibinfo {author} {\bibfnamefont {M.~L.}\ \bibnamefont {Bissell}}, \bibinfo
  {author} {\bibfnamefont {A.~A.}\ \bibnamefont {Breier}}, \bibinfo {author}
  {\bibfnamefont {A.~J.}\ \bibnamefont {Brinson}}, \bibinfo {author}
  {\bibfnamefont {K.}~\bibnamefont {Chrysalidis}}, \bibinfo {author}
  {\bibfnamefont {T.~E.}\ \bibnamefont {Cocolios}}, \bibinfo {author}
  {\bibfnamefont {B.~S.}\ \bibnamefont {Cooper}}, \bibinfo {author}
  {\bibfnamefont {K.~T.}\ \bibnamefont {Flanagan}}, \bibinfo {author}
  {\bibfnamefont {T.~F.}\ \bibnamefont {Giesen}}, \bibinfo {author}
  {\bibfnamefont {R.~P.}\ \bibnamefont {de~Groote}}, \bibinfo {author}
  {\bibfnamefont {S.}~\bibnamefont {Franchoo}}, \bibinfo {author}
  {\bibfnamefont {F.~P.}\ \bibnamefont {Gustafsson}}, \bibinfo {author}
  {\bibfnamefont {T.~A.}\ \bibnamefont {Isaev}}, \bibinfo {author}
  {\bibfnamefont {{\'{A}}.}~\bibnamefont {Koszor{\'{u}}s}}, \bibinfo {author}
  {\bibfnamefont {G.}~\bibnamefont {Neyens}}, \bibinfo {author} {\bibfnamefont
  {H.~A.}\ \bibnamefont {Perrett}}, \bibinfo {author} {\bibfnamefont {C.~M.}\
  \bibnamefont {Ricketts}}, \bibinfo {author} {\bibfnamefont {S.}~\bibnamefont
  {Rothe}}, \bibinfo {author} {\bibfnamefont {L.}~\bibnamefont {Schweikhard}},
  \bibinfo {author} {\bibfnamefont {A.~R.}\ \bibnamefont {Vernon}}, \bibinfo
  {author} {\bibfnamefont {K.~D.~A.}\ \bibnamefont {Wendt}}, \bibinfo {author}
  {\bibfnamefont {F.}~\bibnamefont {Wienholtz}}, \bibinfo {author}
  {\bibfnamefont {S.~G.}\ \bibnamefont {Wilkins}},\ and\ \bibinfo {author}
  {\bibfnamefont {X.~F.}\ \bibnamefont {Yang}},\ }\bibfield  {title} {\bibinfo
  {title} {{Spectroscopy of short-lived radioactive molecules}},\ }\href
  {https://doi.org/10.1038/s41586-020-2299-4} {\bibfield  {journal} {\bibinfo
  {journal} {Nature}\ }\textbf {\bibinfo {volume} {581}},\ \bibinfo {pages}
  {396} (\bibinfo {year} {2020})}\BibitemShut {NoStop}%
\bibitem [{\citenamefont {Hofs\"ass}\ \emph {et~al.}()\citenamefont
  {Hofs\"ass}, \citenamefont {Doppelbauer}, \citenamefont {Wright},
  \citenamefont {Kray}, \citenamefont {Sartakov}, \citenamefont {Pérez-Ríos},
  \citenamefont {Meijer},\ and\ \citenamefont {Truppe}}]{Hofsaess2021}%
  \BibitemOpen
  \bibfield  {author} {\bibinfo {author} {\bibfnamefont {S.}~\bibnamefont
  {Hofs\"ass}}, \bibinfo {author} {\bibfnamefont {M.}~\bibnamefont
  {Doppelbauer}}, \bibinfo {author} {\bibfnamefont {S.}~\bibnamefont {Wright}},
  \bibinfo {author} {\bibfnamefont {S.}~\bibnamefont {Kray}}, \bibinfo {author}
  {\bibfnamefont {B.}~\bibnamefont {Sartakov}}, \bibinfo {author}
  {\bibfnamefont {J.}~\bibnamefont {Pérez-Ríos}}, \bibinfo {author}
  {\bibfnamefont {G.}~\bibnamefont {Meijer}},\ and\ \bibinfo {author}
  {\bibfnamefont {S.}~\bibnamefont {Truppe}},\ }\href@noop {} {}\Eprint
  {https://arxiv.org/abs/2103.15913} {arXiv:2103.15913} \BibitemShut {NoStop}%
\bibitem [{\citenamefont {Grasdijk}\ \emph {et~al.}(2020)\citenamefont
  {Grasdijk}, \citenamefont {Timgren}, \citenamefont {Kastelic}, \citenamefont
  {Wright}, \citenamefont {Lamoreaux}, \citenamefont {DeMille}, \citenamefont
  {Wenz}, \citenamefont {Aitken}, \citenamefont {Zelevinsky}, \citenamefont
  {Winick},\ and\ \citenamefont {Kawall}}]{Grasdijk2020centrex}%
  \BibitemOpen
  \bibfield  {author} {\bibinfo {author} {\bibfnamefont {O.}~\bibnamefont
  {Grasdijk}}, \bibinfo {author} {\bibfnamefont {O.}~\bibnamefont {Timgren}},
  \bibinfo {author} {\bibfnamefont {J.}~\bibnamefont {Kastelic}}, \bibinfo
  {author} {\bibfnamefont {T.}~\bibnamefont {Wright}}, \bibinfo {author}
  {\bibfnamefont {S.}~\bibnamefont {Lamoreaux}}, \bibinfo {author}
  {\bibfnamefont {D.}~\bibnamefont {DeMille}}, \bibinfo {author} {\bibfnamefont
  {K.}~\bibnamefont {Wenz}}, \bibinfo {author} {\bibfnamefont {M.}~\bibnamefont
  {Aitken}}, \bibinfo {author} {\bibfnamefont {T.}~\bibnamefont {Zelevinsky}},
  \bibinfo {author} {\bibfnamefont {T.}~\bibnamefont {Winick}},\ and\ \bibinfo
  {author} {\bibfnamefont {D.}~\bibnamefont {Kawall}},\ }\href@noop {}
  {\bibinfo {title} {Centrex: A new search for time-reversal symmetry violation
  in the $^{205}${Tl} nucleus}} (\bibinfo {year} {2020}),\ \Eprint
  {https://arxiv.org/abs/2010.01451} {arXiv:2010.01451} \BibitemShut {NoStop}%
\bibitem [{\citenamefont {Pilgram}\ \emph {et~al.}()\citenamefont {Pilgram},
  \citenamefont {Jadbabaie}, \citenamefont {Zeng}, \citenamefont {Hutzler},\
  and\ \citenamefont {Steimle}}]{Pilgram2021}%
  \BibitemOpen
  \bibfield  {author} {\bibinfo {author} {\bibfnamefont {N.~H.}\ \bibnamefont
  {Pilgram}}, \bibinfo {author} {\bibfnamefont {A.}~\bibnamefont {Jadbabaie}},
  \bibinfo {author} {\bibfnamefont {Y.}~\bibnamefont {Zeng}}, \bibinfo {author}
  {\bibfnamefont {N.~R.}\ \bibnamefont {Hutzler}},\ and\ \bibinfo {author}
  {\bibfnamefont {T.~C.}\ \bibnamefont {Steimle}},\ }\href@noop {} {}\Eprint
  {https://arxiv.org/abs/2104.11769} {arXiv:2104.11769} \BibitemShut {NoStop}%
\bibitem [{\citenamefont {Steimle}\ \emph {et~al.}(2011)\citenamefont
  {Steimle}, \citenamefont {Frey}, \citenamefont {Le}, \citenamefont {DeMille},
  \citenamefont {Rahmlow},\ and\ \citenamefont {Linton}}]{Steimle2011}%
  \BibitemOpen
  \bibfield  {author} {\bibinfo {author} {\bibfnamefont {T.~C.}\ \bibnamefont
  {Steimle}}, \bibinfo {author} {\bibfnamefont {S.}~\bibnamefont {Frey}},
  \bibinfo {author} {\bibfnamefont {A.}~\bibnamefont {Le}}, \bibinfo {author}
  {\bibfnamefont {D.}~\bibnamefont {DeMille}}, \bibinfo {author} {\bibfnamefont
  {D.~A.}\ \bibnamefont {Rahmlow}},\ and\ \bibinfo {author} {\bibfnamefont
  {C.}~\bibnamefont {Linton}},\ }\bibfield  {title} {\bibinfo {title}
  {Molecular-beam optical {Stark} and {Zeeman} study of the $a$
  ${}^{2}\ensuremath{\Pi}$--$x$ ${}^{2}{\ensuremath{\Sigma}}^{+}$ (0,0) band
  system of {BaF}},\ }\href {https://doi.org/10.1103/PhysRevA.84.012508}
  {\bibfield  {journal} {\bibinfo  {journal} {Phys. Rev. A}\ }\textbf {\bibinfo
  {volume} {84}},\ \bibinfo {pages} {012508} (\bibinfo {year}
  {2011})}\BibitemShut {NoStop}%
\bibitem [{\citenamefont {{Di Rosa}}(2004)}]{DiRosa2004}%
  \BibitemOpen
  \bibfield  {author} {\bibinfo {author} {\bibfnamefont {M.~D.}\ \bibnamefont
  {{Di Rosa}}},\ }\bibfield  {title} {\bibinfo {title} {{Laser-cooling
  molecules}},\ }\href {https://doi.org/10.1140/epjd/e2004-00167-2} {\bibfield
  {journal} {\bibinfo  {journal} {The European Physical Journal D - Atomic,
  Molecular, Optical and Plasma Physics}\ }\textbf {\bibinfo {volume} {31}},\
  \bibinfo {pages} {395} (\bibinfo {year} {2004})}\BibitemShut {NoStop}%
\bibitem [{\citenamefont {Chen}\ \emph {et~al.}(2016)\citenamefont {Chen},
  \citenamefont {Bu},\ and\ \citenamefont {Yan}}]{Chen2016}%
  \BibitemOpen
  \bibfield  {author} {\bibinfo {author} {\bibfnamefont {T.}~\bibnamefont
  {Chen}}, \bibinfo {author} {\bibfnamefont {W.}~\bibnamefont {Bu}},\ and\
  \bibinfo {author} {\bibfnamefont {B.}~\bibnamefont {Yan}},\ }\bibfield
  {title} {\bibinfo {title} {Structure, branching ratios, and a laser-cooling
  scheme for the $^{138}\mathrm{BaF}$ molecule},\ }\href
  {https://doi.org/10.1103/PhysRevA.94.063415} {\bibfield  {journal} {\bibinfo
  {journal} {Phys. Rev. A}\ }\textbf {\bibinfo {volume} {94}},\ \bibinfo
  {pages} {063415} (\bibinfo {year} {2016})}\BibitemShut {NoStop}%
\bibitem [{\citenamefont {Stuhl}\ \emph {et~al.}(2008)\citenamefont {Stuhl},
  \citenamefont {Sawyer}, \citenamefont {Wang},\ and\ \citenamefont
  {Ye}}]{Stuhl2008}%
  \BibitemOpen
  \bibfield  {author} {\bibinfo {author} {\bibfnamefont {B.~K.}\ \bibnamefont
  {Stuhl}}, \bibinfo {author} {\bibfnamefont {B.~C.}\ \bibnamefont {Sawyer}},
  \bibinfo {author} {\bibfnamefont {D.}~\bibnamefont {Wang}},\ and\ \bibinfo
  {author} {\bibfnamefont {J.}~\bibnamefont {Ye}},\ }\bibfield  {title}
  {\bibinfo {title} {Magneto-optical trap for polar molecules},\ }\href
  {https://doi.org/10.1103/PhysRevLett.101.243002} {\bibfield  {journal}
  {\bibinfo  {journal} {Phys. Rev. Lett.}\ }\textbf {\bibinfo {volume} {101}},\
  \bibinfo {pages} {243002} (\bibinfo {year} {2008})}\BibitemShut {NoStop}%
\bibitem [{app()}]{appendix}%
  \BibitemOpen
  \href@noop {} {\bibinfo {title} {See appendix for more details.}}\BibitemShut
  {Stop}%
\bibitem [{Note1()}]{Note1}%
  \BibitemOpen
  \bibinfo {note} {As the interaction strength between rotation $\protect
  \mathbf {N}$ and intermediate angular momentum $\protect \mathbf {G}$ is
  comparable to the hyperfine coupling of the second nuclear spin $\protect
  \mathbf {I}_\protect \mathrm {F}$, the effective Hamiltonian mixes different
  $F_1$ and cause a violation of the selection rule $\Delta F_1 = 0,\pm 1$,
  rendering two transitions only approximately forbidden.}\BibitemShut {Stop}%
\bibitem [{\citenamefont {Rogers}\ \emph {et~al.}(2011)\citenamefont {Rogers},
  \citenamefont {Carini}, \citenamefont {Pechkis},\ and\ \citenamefont
  {Gould}}]{Rogers2011}%
  \BibitemOpen
  \bibfield  {author} {\bibinfo {author} {\bibfnamefont {C.~E.}\ \bibnamefont
  {Rogers}}, \bibinfo {author} {\bibfnamefont {J.~L.}\ \bibnamefont {Carini}},
  \bibinfo {author} {\bibfnamefont {J.~A.}\ \bibnamefont {Pechkis}},\ and\
  \bibinfo {author} {\bibfnamefont {P.~L.}\ \bibnamefont {Gould}},\ }\bibfield
  {title} {\bibinfo {title} {Creation of arbitrary time-sequenced line spectra
  with an electro-optic phase modulator},\ }\href
  {https://doi.org/10.1063/1.3611005} {\bibfield  {journal} {\bibinfo
  {journal} {Review of Scientific Instruments}\ }\textbf {\bibinfo {volume}
  {82}},\ \bibinfo {pages} {073107} (\bibinfo {year} {2011})}\BibitemShut
  {NoStop}%
\bibitem [{\citenamefont {Holland}\ \emph {et~al.}(2021)\citenamefont
  {Holland}, \citenamefont {Lu},\ and\ \citenamefont {Cheuk}}]{Holland2021}%
  \BibitemOpen
  \bibfield  {author} {\bibinfo {author} {\bibfnamefont {C.~M.}\ \bibnamefont
  {Holland}}, \bibinfo {author} {\bibfnamefont {Y.}~\bibnamefont {Lu}},\ and\
  \bibinfo {author} {\bibfnamefont {L.~W.}\ \bibnamefont {Cheuk}},\ }\bibfield
  {title} {\bibinfo {title} {Synthesizing optical spectra using
  computer-generated holography techniques},\ }\href
  {https://doi.org/10.1088/1367-2630/abe973} {\bibfield  {journal} {\bibinfo
  {journal} {New Journal of Physics}\ }\textbf {\bibinfo {volume} {23}},\
  \bibinfo {pages} {033028} (\bibinfo {year} {2021})}\BibitemShut {NoStop}%
\bibitem [{\citenamefont {Devlin}\ and\ \citenamefont
  {Tarbutt}(2018)}]{Devlin2018}%
  \BibitemOpen
  \bibfield  {author} {\bibinfo {author} {\bibfnamefont {J.~A.}\ \bibnamefont
  {Devlin}}\ and\ \bibinfo {author} {\bibfnamefont {M.~R.}\ \bibnamefont
  {Tarbutt}},\ }\bibfield  {title} {\bibinfo {title} {Laser cooling and
  magneto-optical trapping of molecules analyzed using optical {Bloch}
  equations and the {Fokker-Planck-Kramers} equation},\ }\href
  {https://doi.org/10.1103/PhysRevA.98.063415} {\bibfield  {journal} {\bibinfo
  {journal} {Phys. Rev. A}\ }\textbf {\bibinfo {volume} {98}},\ \bibinfo
  {pages} {063415} (\bibinfo {year} {2018})}\BibitemShut {NoStop}%
\bibitem [{\citenamefont {Kogel}\ and\ \citenamefont
  {Langen}(2021)}]{Kogel2021code}%
  \BibitemOpen
  \bibfield  {author} {\bibinfo {author} {\bibfnamefont {F.}~\bibnamefont
  {Kogel}}\ and\ \bibinfo {author} {\bibfnamefont {T.}~\bibnamefont {Langen}},\
  }\href@noop {} {\bibfield  {journal} {\bibinfo  {journal} {in preparation}\ }
  (\bibinfo {year} {2021})}\BibitemShut {NoStop}%
\bibitem [{\citenamefont {Barry}(2013)}]{BarryPhD}%
  \BibitemOpen
  \bibfield  {author} {\bibinfo {author} {\bibfnamefont {J.~F.}\ \bibnamefont
  {Barry}},\ }\emph {\bibinfo {title} {Laser cooling and slowing of a diatomic
  molecule}},\ \href@noop {} {\bibinfo {type} {dissertation}},\ \bibinfo
  {school} {Yale University} (\bibinfo {year} {2013})\BibitemShut {NoStop}%
\bibitem [{\citenamefont {McNally}(2021)}]{McNallyPhD}%
  \BibitemOpen
  \bibfield  {author} {\bibinfo {author} {\bibfnamefont {R.}~\bibnamefont
  {McNally}},\ }\emph {\bibinfo {title} {Laser cooling of BaH molecules, and
  new ideas for the detection of dark matter}},\ \href@noop {} {\bibinfo {type}
  {dissertation}},\ \bibinfo  {school} {Columbia University} (\bibinfo {year}
  {2021})\BibitemShut {NoStop}%
\bibitem [{\citenamefont {Tarbutt}(2015)}]{Tarbutt2015}%
  \BibitemOpen
  \bibfield  {author} {\bibinfo {author} {\bibfnamefont {M.~R.}\ \bibnamefont
  {Tarbutt}},\ }\bibfield  {title} {\bibinfo {title} {Magneto-optical trapping
  forces for atoms and molecules with complex level structures},\ }\href@noop
  {} {\bibfield  {journal} {\bibinfo  {journal} {New Journal of Physics}\
  }\textbf {\bibinfo {volume} {17}},\ \bibinfo {pages} {015007} (\bibinfo
  {year} {2015})}\BibitemShut {NoStop}%
\bibitem [{\citenamefont {Alauze}\ \emph {et~al.}()\citenamefont {Alauze},
  \citenamefont {Lim}, \citenamefont {Trigatzis}, \citenamefont {Swarbrick},
  \citenamefont {Fitch}, \citenamefont {Sauer},\ and\ \citenamefont
  {Tarbutt}}]{Alauze2021}%
  \BibitemOpen
  \bibfield  {author} {\bibinfo {author} {\bibfnamefont {X.}~\bibnamefont
  {Alauze}}, \bibinfo {author} {\bibfnamefont {J.}~\bibnamefont {Lim}},
  \bibinfo {author} {\bibfnamefont {M.~A.}\ \bibnamefont {Trigatzis}}, \bibinfo
  {author} {\bibfnamefont {S.}~\bibnamefont {Swarbrick}}, \bibinfo {author}
  {\bibfnamefont {N.~J.}\ \bibnamefont {Fitch}}, \bibinfo {author}
  {\bibfnamefont {B.~E.}\ \bibnamefont {Sauer}},\ and\ \bibinfo {author}
  {\bibfnamefont {M.~R.}\ \bibnamefont {Tarbutt}},\ }\href@noop {} {}\Eprint
  {https://arxiv.org/abs/2104.06194} {arXiv:2104.06194} \BibitemShut {NoStop}%
\bibitem [{\citenamefont {S\"oding}\ \emph {et~al.}(1997)\citenamefont
  {S\"oding}, \citenamefont {Grimm}, \citenamefont {Ovchinnikov}, \citenamefont
  {Bouyer},\ and\ \citenamefont {Salomon}}]{Soeding1997}%
  \BibitemOpen
  \bibfield  {author} {\bibinfo {author} {\bibfnamefont {J.}~\bibnamefont
  {S\"oding}}, \bibinfo {author} {\bibfnamefont {R.}~\bibnamefont {Grimm}},
  \bibinfo {author} {\bibfnamefont {Y.~B.}\ \bibnamefont {Ovchinnikov}},
  \bibinfo {author} {\bibfnamefont {P.}~\bibnamefont {Bouyer}},\ and\ \bibinfo
  {author} {\bibfnamefont {C.}~\bibnamefont {Salomon}},\ }\bibfield  {title}
  {\bibinfo {title} {Short-distance atomic beam deceleration with a stimulated
  light force},\ }\href {https://doi.org/10.1103/PhysRevLett.78.1420}
  {\bibfield  {journal} {\bibinfo  {journal} {Phys. Rev. Lett.}\ }\textbf
  {\bibinfo {volume} {78}},\ \bibinfo {pages} {1420} (\bibinfo {year}
  {1997})}\BibitemShut {NoStop}%
\bibitem [{\citenamefont {Chieda}\ and\ \citenamefont
  {Eyler}(2012)}]{Chieda2012}%
  \BibitemOpen
  \bibfield  {author} {\bibinfo {author} {\bibfnamefont {M.~A.}\ \bibnamefont
  {Chieda}}\ and\ \bibinfo {author} {\bibfnamefont {E.~E.}\ \bibnamefont
  {Eyler}},\ }\bibfield  {title} {\bibinfo {title} {{Bichromatic slowing of
  metastable helium}},\ }\href {https://doi.org/10.1103/PhysRevA.86.053415}
  {\bibfield  {journal} {\bibinfo  {journal} {Phys. Rev. A}\ }\textbf {\bibinfo
  {volume} {86}},\ \bibinfo {pages} {53415} (\bibinfo {year}
  {2012})}\BibitemShut {NoStop}%
\bibitem [{\citenamefont {Kozyryev}\ \emph {et~al.}(2018)\citenamefont
  {Kozyryev}, \citenamefont {Baum}, \citenamefont {Aldridge}, \citenamefont
  {Yu}, \citenamefont {Eyler},\ and\ \citenamefont
  {Doyle}}]{KozyryevBichromatic2018}%
  \BibitemOpen
  \bibfield  {author} {\bibinfo {author} {\bibfnamefont {I.}~\bibnamefont
  {Kozyryev}}, \bibinfo {author} {\bibfnamefont {L.}~\bibnamefont {Baum}},
  \bibinfo {author} {\bibfnamefont {L.}~\bibnamefont {Aldridge}}, \bibinfo
  {author} {\bibfnamefont {P.}~\bibnamefont {Yu}}, \bibinfo {author}
  {\bibfnamefont {E.~E.}\ \bibnamefont {Eyler}},\ and\ \bibinfo {author}
  {\bibfnamefont {J.~M.}\ \bibnamefont {Doyle}},\ }\bibfield  {title} {\bibinfo
  {title} {{Coherent Bichromatic Force Deflection of Molecules}},\ }\href
  {https://doi.org/10.1103/PhysRevLett.120.063205} {\bibfield  {journal}
  {\bibinfo  {journal} {Phys. Rev. Lett.}\ }\textbf {\bibinfo {volume} {120}},\
  \bibinfo {pages} {63205} (\bibinfo {year} {2018})}\BibitemShut {NoStop}%
\bibitem [{\citenamefont {Galica}\ \emph {et~al.}(2018)\citenamefont {Galica},
  \citenamefont {Aldridge}, \citenamefont {McCarron}, \citenamefont {Eyler},\
  and\ \citenamefont {Gould}}]{GalicaDeflection2018}%
  \BibitemOpen
  \bibfield  {author} {\bibinfo {author} {\bibfnamefont {S.~E.}\ \bibnamefont
  {Galica}}, \bibinfo {author} {\bibfnamefont {L.}~\bibnamefont {Aldridge}},
  \bibinfo {author} {\bibfnamefont {D.~J.}\ \bibnamefont {McCarron}}, \bibinfo
  {author} {\bibfnamefont {E.~E.}\ \bibnamefont {Eyler}},\ and\ \bibinfo
  {author} {\bibfnamefont {P.~L.}\ \bibnamefont {Gould}},\ }\bibfield  {title}
  {\bibinfo {title} {{Deflection of a molecular beam using the bichromatic
  stimulated force}},\ }\href {https://doi.org/10.1103/PhysRevA.98.023408}
  {\bibfield  {journal} {\bibinfo  {journal} {Phys. Rev. A}\ }\textbf {\bibinfo
  {volume} {98}},\ \bibinfo {pages} {23408} (\bibinfo {year}
  {2018})}\BibitemShut {NoStop}%
\bibitem [{Note2()}]{Note2}%
  \BibitemOpen
  \bibinfo {note} {We note that, while the simplified level scheme shown in
  Fig.~\ref {fig:bichromatic}a is similar to a scheme that was recently
  proposed for the realization of large molasses-like bichromatic cooling
  forces~\cite {Wenz2020a}, such forces can not be realized in {${}^{137}$BaF}\
  efficiently due to the various loss channels discussed in the main
  text.}\BibitemShut {Stop}%
\bibitem [{\citenamefont {Aldridge}\ \emph {et~al.}(2016)\citenamefont
  {Aldridge}, \citenamefont {Galica},\ and\ \citenamefont
  {Eyler}}]{Aldridge2016}%
  \BibitemOpen
  \bibfield  {author} {\bibinfo {author} {\bibfnamefont {L.}~\bibnamefont
  {Aldridge}}, \bibinfo {author} {\bibfnamefont {S.~E.}\ \bibnamefont
  {Galica}},\ and\ \bibinfo {author} {\bibfnamefont {E.~E.}\ \bibnamefont
  {Eyler}},\ }\bibfield  {title} {\bibinfo {title} {{Simulations of the
  bichromatic force in multilevel systems}},\ }\href
  {https://doi.org/10.1103/PhysRevA.93.013419} {\bibfield  {journal} {\bibinfo
  {journal} {Physical Review A}\ }\textbf {\bibinfo {volume} {93}},\ \bibinfo
  {pages} {1} (\bibinfo {year} {2016})},\ \Eprint
  {https://arxiv.org/abs/1509.05350} {arXiv:1509.05350} \BibitemShut {NoStop}%
\bibitem [{Note3()}]{Note3}%
  \BibitemOpen
  \bibinfo {note} {Note that while the BCF light field addresses all ground
  states in $G=2$, transition 7, which shares its groundstate with the direct
  repumper on transition 1, is dipole forbidden. This avoids any interference
  between the direct repumper and the BCF light field.}\BibitemShut {Stop}%
\bibitem [{\citenamefont {Norrgard}\ \emph {et~al.}(2019)\citenamefont
  {Norrgard}, \citenamefont {Barker}, \citenamefont {Eckel}, \citenamefont
  {Fedchak}, \citenamefont {Klimov},\ and\ \citenamefont
  {Scherschligt}}]{Norrgard2019}%
  \BibitemOpen
  \bibfield  {author} {\bibinfo {author} {\bibfnamefont {E.~B.}\ \bibnamefont
  {Norrgard}}, \bibinfo {author} {\bibfnamefont {D.~S.}\ \bibnamefont
  {Barker}}, \bibinfo {author} {\bibfnamefont {S.}~\bibnamefont {Eckel}},
  \bibinfo {author} {\bibfnamefont {J.~A.}\ \bibnamefont {Fedchak}}, \bibinfo
  {author} {\bibfnamefont {N.~N.}\ \bibnamefont {Klimov}},\ and\ \bibinfo
  {author} {\bibfnamefont {J.}~\bibnamefont {Scherschligt}},\ }\bibfield
  {title} {\bibinfo {title} {{Nuclear-spin dependent parity violation in
  optically trapped polyatomic molecules}},\ }\href
  {https://doi.org/10.1038/s42005-019-0181-1} {\bibfield  {journal} {\bibinfo
  {journal} {Communications Physics}\ }\textbf {\bibinfo {volume} {2}},\
  \bibinfo {pages} {77} (\bibinfo {year} {2019})}\BibitemShut {NoStop}%
\bibitem [{\citenamefont {Dzuba}\ \emph {et~al.}(2017)\citenamefont {Dzuba},
  \citenamefont {Flambaum},\ and\ \citenamefont {Stadnik}}]{Dzuba2017}%
  \BibitemOpen
  \bibfield  {author} {\bibinfo {author} {\bibfnamefont {V.~A.}\ \bibnamefont
  {Dzuba}}, \bibinfo {author} {\bibfnamefont {V.~V.}\ \bibnamefont
  {Flambaum}},\ and\ \bibinfo {author} {\bibfnamefont {Y.~V.}\ \bibnamefont
  {Stadnik}},\ }\bibfield  {title} {\bibinfo {title} {{Probing Low-Mass Vector
  Bosons with Parity Nonconservation and Nuclear Anapole Moment Measurements in
  Atoms and Molecules}},\ }\href
  {https://doi.org/10.1103/PhysRevLett.119.223201} {\bibfield  {journal}
  {\bibinfo  {journal} {Phys. Rev. Lett.}\ }\textbf {\bibinfo {volume} {119}},\
  \bibinfo {pages} {223201} (\bibinfo {year} {2017})}\BibitemShut {NoStop}%
\bibitem [{\citenamefont {Altunta\ifmmode~\mbox{\c{s}}\else \c{s}\fi{}}\ \emph
  {et~al.}(2018)\citenamefont {Altunta\ifmmode~\mbox{\c{s}}\else \c{s}\fi{}},
  \citenamefont {Ammon}, \citenamefont {Cahn},\ and\ \citenamefont
  {DeMille}}]{AltuntasPRA}%
  \BibitemOpen
  \bibfield  {author} {\bibinfo {author} {\bibfnamefont {E.}~\bibnamefont
  {Altunta\ifmmode~\mbox{\c{s}}\else \c{s}\fi{}}}, \bibinfo {author}
  {\bibfnamefont {J.}~\bibnamefont {Ammon}}, \bibinfo {author} {\bibfnamefont
  {S.~B.}\ \bibnamefont {Cahn}},\ and\ \bibinfo {author} {\bibfnamefont
  {D.}~\bibnamefont {DeMille}},\ }\bibfield  {title} {\bibinfo {title}
  {Measuring nuclear-spin-dependent parity violation with molecules:
  Experimental methods and analysis of systematic errors},\ }\href
  {https://doi.org/10.1103/PhysRevA.97.042101} {\bibfield  {journal} {\bibinfo
  {journal} {Phys. Rev. A}\ }\textbf {\bibinfo {volume} {97}},\ \bibinfo
  {pages} {042101} (\bibinfo {year} {2018})}\BibitemShut {NoStop}%
\bibitem [{\citenamefont {Rahmlow}(2010)}]{RahmlowPhD}%
  \BibitemOpen
  \bibfield  {author} {\bibinfo {author} {\bibfnamefont {D.}~\bibnamefont
  {Rahmlow}},\ }\emph {\bibinfo {title} {Towards a measurement of parity
  nonconservation in diatomic molecules}},\ \href@noop {} {\bibinfo {type}
  {dissertation}},\ \bibinfo  {school} {Yale University} (\bibinfo {year}
  {2010})\BibitemShut {NoStop}%
\bibitem [{\citenamefont {Shaw}\ and\ \citenamefont
  {McCarron}(2020)}]{Shaw2020}%
  \BibitemOpen
  \bibfield  {author} {\bibinfo {author} {\bibfnamefont {J.~C.}\ \bibnamefont
  {Shaw}}\ and\ \bibinfo {author} {\bibfnamefont {D.~J.}\ \bibnamefont
  {McCarron}},\ }\bibfield  {title} {\bibinfo {title} {Bright, continuous beams
  of cold free radicals},\ }\href {https://doi.org/10.1103/PhysRevA.102.041302}
  {\bibfield  {journal} {\bibinfo  {journal} {Phys. Rev. A}\ }\textbf {\bibinfo
  {volume} {102}},\ \bibinfo {pages} {041302} (\bibinfo {year}
  {2020})}\BibitemShut {NoStop}%
\bibitem [{\citenamefont {Augenbraun}\ \emph {et~al.}(2020)\citenamefont
  {Augenbraun}, \citenamefont {Lasner}, \citenamefont {Frenett}, \citenamefont
  {Sawaoka}, \citenamefont {Miller}, \citenamefont {Steimle},\ and\
  \citenamefont {Doyle}}]{Augenbraun2020}%
  \BibitemOpen
  \bibfield  {author} {\bibinfo {author} {\bibfnamefont {B.~L.}\ \bibnamefont
  {Augenbraun}}, \bibinfo {author} {\bibfnamefont {Z.~D.}\ \bibnamefont
  {Lasner}}, \bibinfo {author} {\bibfnamefont {A.}~\bibnamefont {Frenett}},
  \bibinfo {author} {\bibfnamefont {H.}~\bibnamefont {Sawaoka}}, \bibinfo
  {author} {\bibfnamefont {C.}~\bibnamefont {Miller}}, \bibinfo {author}
  {\bibfnamefont {T.~C.}\ \bibnamefont {Steimle}},\ and\ \bibinfo {author}
  {\bibfnamefont {J.~M.}\ \bibnamefont {Doyle}},\ }\bibfield  {title} {\bibinfo
  {title} {Laser-cooled polyatomic molecules for improved electron electric
  dipole moment searches},\ }\href {https://doi.org/10.1088/1367-2630/ab687b}
  {\bibfield  {journal} {\bibinfo  {journal} {New Journal of Physics}\ }\textbf
  {\bibinfo {volume} {22}},\ \bibinfo {pages} {022003} (\bibinfo {year}
  {2020})}\BibitemShut {NoStop}%
\bibitem [{\citenamefont {Jadbabaie}\ \emph {et~al.}(2020)\citenamefont
  {Jadbabaie}, \citenamefont {Pilgram}, \citenamefont {Klos}, \citenamefont
  {Kotochigova},\ and\ \citenamefont {Hutzler}}]{Jadbabaie2020}%
  \BibitemOpen
  \bibfield  {author} {\bibinfo {author} {\bibfnamefont {A.}~\bibnamefont
  {Jadbabaie}}, \bibinfo {author} {\bibfnamefont {N.~H.}\ \bibnamefont
  {Pilgram}}, \bibinfo {author} {\bibfnamefont {J.}~\bibnamefont {Klos}},
  \bibinfo {author} {\bibfnamefont {S.}~\bibnamefont {Kotochigova}},\ and\
  \bibinfo {author} {\bibfnamefont {N.~R.}\ \bibnamefont {Hutzler}},\
  }\bibfield  {title} {\bibinfo {title} {Enhanced molecular yield from a
  cryogenic buffer gas beam source via excited state chemistry},\ }\href
  {https://doi.org/10.1088/1367-2630/ab6eae} {\bibfield  {journal} {\bibinfo
  {journal} {New Journal of Physics}\ }\textbf {\bibinfo {volume} {22}},\
  \bibinfo {pages} {022002} (\bibinfo {year} {2020})}\BibitemShut {NoStop}%
\bibitem [{\citenamefont {Fitch}\ \emph {et~al.}()\citenamefont {Fitch},
  \citenamefont {Lim}, \citenamefont {Hinds}, \citenamefont {Sauer},\ and\
  \citenamefont {Tarbutt}}]{Fitch2020methods}%
  \BibitemOpen
  \bibfield  {author} {\bibinfo {author} {\bibfnamefont {N.~J.}\ \bibnamefont
  {Fitch}}, \bibinfo {author} {\bibfnamefont {J.}~\bibnamefont {Lim}}, \bibinfo
  {author} {\bibfnamefont {E.~A.}\ \bibnamefont {Hinds}}, \bibinfo {author}
  {\bibfnamefont {B.~E.}\ \bibnamefont {Sauer}},\ and\ \bibinfo {author}
  {\bibfnamefont {M.~R.}\ \bibnamefont {Tarbutt}},\ }\href@noop {} {}\Eprint
  {https://arxiv.org/abs/2009.00346} {arXiv:2009.00346} \BibitemShut {NoStop}%
\bibitem [{Note4()}]{Note4}%
  \BibitemOpen
  \bibinfo {note} {In this case, a minimum of $48$ more ground states need to
  be taken into account leading to a reduced scattering rate $R_\protect
  \mathrm {sc}=\Gamma /7$}\BibitemShut {NoStop}%
\bibitem [{\citenamefont {Yeo}\ \emph {et~al.}(2015)\citenamefont {Yeo},
  \citenamefont {Hummon}, \citenamefont {Collopy}, \citenamefont {Yan},
  \citenamefont {Hemmerling}, \citenamefont {Chae}, \citenamefont {Doyle},\
  and\ \citenamefont {Ye}}]{Yeo2015}%
  \BibitemOpen
  \bibfield  {author} {\bibinfo {author} {\bibfnamefont {M.}~\bibnamefont
  {Yeo}}, \bibinfo {author} {\bibfnamefont {M.~T.}\ \bibnamefont {Hummon}},
  \bibinfo {author} {\bibfnamefont {A.~L.}\ \bibnamefont {Collopy}}, \bibinfo
  {author} {\bibfnamefont {B.}~\bibnamefont {Yan}}, \bibinfo {author}
  {\bibfnamefont {B.}~\bibnamefont {Hemmerling}}, \bibinfo {author}
  {\bibfnamefont {E.}~\bibnamefont {Chae}}, \bibinfo {author} {\bibfnamefont
  {J.~M.}\ \bibnamefont {Doyle}},\ and\ \bibinfo {author} {\bibfnamefont
  {J.}~\bibnamefont {Ye}},\ }\bibfield  {title} {\bibinfo {title} {Rotational
  state microwave mixing for laser cooling of complex diatomic molecules},\
  }\href {https://doi.org/10.1103/PhysRevLett.114.223003} {\bibfield  {journal}
  {\bibinfo  {journal} {Phys. Rev. Lett.}\ }\textbf {\bibinfo {volume} {114}},\
  \bibinfo {pages} {223003} (\bibinfo {year} {2015})}\BibitemShut {NoStop}%
\bibitem [{\citenamefont {Hao}\ \emph {et~al.}(2019)\citenamefont {Hao},
  \citenamefont {Pašteka}, \citenamefont {Visscher}, \citenamefont {Aggarwal},
  \citenamefont {Bethlem}, \citenamefont {Boeschoten}, \citenamefont
  {Borschevsky}, \citenamefont {Denis}, \citenamefont {Esajas}, \citenamefont
  {Hoekstra}, \citenamefont {Jungmann}, \citenamefont {Marshall}, \citenamefont
  {Meijknecht}, \citenamefont {Mooij}, \citenamefont {Timmermans},
  \citenamefont {Touwen}, \citenamefont {Ubachs}, \citenamefont {Willmann},
  \citenamefont {Yin},\ and\ \citenamefont {Zapara}}]{Hao2019}%
  \BibitemOpen
  \bibfield  {author} {\bibinfo {author} {\bibfnamefont {Y.}~\bibnamefont
  {Hao}}, \bibinfo {author} {\bibfnamefont {L.~F.}\ \bibnamefont {Pašteka}},
  \bibinfo {author} {\bibfnamefont {L.}~\bibnamefont {Visscher}}, \bibinfo
  {author} {\bibfnamefont {P.}~\bibnamefont {Aggarwal}}, \bibinfo {author}
  {\bibfnamefont {H.~L.}\ \bibnamefont {Bethlem}}, \bibinfo {author}
  {\bibfnamefont {A.}~\bibnamefont {Boeschoten}}, \bibinfo {author}
  {\bibfnamefont {A.}~\bibnamefont {Borschevsky}}, \bibinfo {author}
  {\bibfnamefont {M.}~\bibnamefont {Denis}}, \bibinfo {author} {\bibfnamefont
  {K.}~\bibnamefont {Esajas}}, \bibinfo {author} {\bibfnamefont
  {S.}~\bibnamefont {Hoekstra}}, \bibinfo {author} {\bibfnamefont
  {K.}~\bibnamefont {Jungmann}}, \bibinfo {author} {\bibfnamefont {V.~R.}\
  \bibnamefont {Marshall}}, \bibinfo {author} {\bibfnamefont {T.~B.}\
  \bibnamefont {Meijknecht}}, \bibinfo {author} {\bibfnamefont {M.~C.}\
  \bibnamefont {Mooij}}, \bibinfo {author} {\bibfnamefont {R.~G.~E.}\
  \bibnamefont {Timmermans}}, \bibinfo {author} {\bibfnamefont
  {A.}~\bibnamefont {Touwen}}, \bibinfo {author} {\bibfnamefont
  {W.}~\bibnamefont {Ubachs}}, \bibinfo {author} {\bibfnamefont
  {L.}~\bibnamefont {Willmann}}, \bibinfo {author} {\bibfnamefont
  {Y.}~\bibnamefont {Yin}},\ and\ \bibinfo {author} {\bibfnamefont
  {A.}~\bibnamefont {Zapara}},\ }\bibfield  {title} {\bibinfo {title} {High
  accuracy theoretical investigations of {CaF}, {SrF}, and {BaF} and
  implications for laser-cooling},\ }\href@noop {} {\bibfield  {journal}
  {\bibinfo  {journal} {The Journal of Chemical Physics}\ }\textbf {\bibinfo
  {volume} {151}},\ \bibinfo {pages} {034302} (\bibinfo {year}
  {2019})}\BibitemShut {NoStop}%
\bibitem [{\citenamefont {Balakrishnan}(2016)}]{Balakrishnan2016}%
  \BibitemOpen
  \bibfield  {author} {\bibinfo {author} {\bibfnamefont {N.}~\bibnamefont
  {Balakrishnan}},\ }\bibfield  {title} {\bibinfo {title} {Perspective:
  Ultracold molecules and the dawn of cold controlled chemistry},\ }\href
  {https://doi.org/10.1063/1.4964096} {\bibfield  {journal} {\bibinfo
  {journal} {The Journal of Chemical Physics}\ }\textbf {\bibinfo {volume}
  {145}},\ \bibinfo {pages} {150901} (\bibinfo {year} {2016})}\BibitemShut
  {NoStop}%
\bibitem [{\citenamefont {Tomza}(2015)}]{Tomza2015}%
  \BibitemOpen
  \bibfield  {author} {\bibinfo {author} {\bibfnamefont {M.}~\bibnamefont
  {Tomza}},\ }\bibfield  {title} {\bibinfo {title} {Energetics and control of
  ultracold isotope-exchange reactions between heteronuclear dimers in external
  fields},\ }\href {https://doi.org/10.1103/PhysRevLett.115.063201} {\bibfield
  {journal} {\bibinfo  {journal} {Phys. Rev. Lett.}\ }\textbf {\bibinfo
  {volume} {115}},\ \bibinfo {pages} {063201} (\bibinfo {year}
  {2015})}\BibitemShut {NoStop}%
\bibitem [{\citenamefont {Brown}\ and\ \citenamefont
  {Carrington}(2003)}]{Brown2003}%
  \BibitemOpen
  \bibfield  {author} {\bibinfo {author} {\bibfnamefont {J.}~\bibnamefont
  {Brown}}\ and\ \bibinfo {author} {\bibfnamefont {A.}~\bibnamefont
  {Carrington}},\ }\href {https://books.google.de/books?id=BcZHngEACAAJ} {\emph
  {\bibinfo {title} {Rotational Spectroscopy of Diatomic Molecules}}},\
  Cambridge molecular science series\ (\bibinfo  {publisher} {Cambridge
  University Press},\ \bibinfo {year} {2003})\BibitemShut {NoStop}%
\bibitem [{\citenamefont {Wenz}\ \emph {et~al.}(2020)\citenamefont {Wenz},
  \citenamefont {Kozyryev}, \citenamefont {McNally}, \citenamefont {Aldridge},\
  and\ \citenamefont {Zelevinsky}}]{Wenz2020a}%
  \BibitemOpen
  \bibfield  {author} {\bibinfo {author} {\bibfnamefont {K.}~\bibnamefont
  {Wenz}}, \bibinfo {author} {\bibfnamefont {I.}~\bibnamefont {Kozyryev}},
  \bibinfo {author} {\bibfnamefont {R.~L.}\ \bibnamefont {McNally}}, \bibinfo
  {author} {\bibfnamefont {L.}~\bibnamefont {Aldridge}},\ and\ \bibinfo
  {author} {\bibfnamefont {T.}~\bibnamefont {Zelevinsky}},\ }\bibfield  {title}
  {\bibinfo {title} {Large molasses-like cooling forces for molecules using
  polychromatic optical fields: A theoretical description},\ }\href
  {https://doi.org/10.1103/PhysRevResearch.2.043377} {\bibfield  {journal}
  {\bibinfo  {journal} {Phys. Rev. Research}\ }\textbf {\bibinfo {volume}
  {2}},\ \bibinfo {pages} {043377} (\bibinfo {year} {2020})}\BibitemShut
  {NoStop}%
\end{thebibliography}%
\end{document}